\documentclass[prd,twocolumn,showpacs,superscriptaddress,nofootinbib,floatfix,showkeys,10pt]{revtex4-2}
\usepackage{bm}
\usepackage{times}
\usepackage{braket}
\usepackage{amsfonts,amssymb,stmaryrd,latexsym,amsmath}
\usepackage[usenames,dvipsnames]{color}
\usepackage{slashed}
\usepackage{hyperref}
\usepackage{subfigure}
\usepackage{graphicx}
\usepackage{slashed}
\usepackage{orcidlink}
\usepackage{multirow}
\usepackage{appendix}
\usepackage{dashrule}
\usepackage{float}
\allowdisplaybreaks[1]
\usepackage{verbatim}

\newcommand{\Tr}{\text{Tr}}
\newcommand{\etal}{\textit{et al}. }

\def\DD{{DD^{*}}}
\def\DDbar{{D\bar{D}^{*}/\bar{D}D^*}}

\definecolor{nicered}{rgb}{0.7,0.1,0.1}
\definecolor{nicegreen}{rgb}{0.1,0.5,0.1}
\definecolor{emph}{rgb}{1,0,0}
\definecolor{doub}{rgb}{0.7,0.2,1.0}
\definecolor{navyblue}{RGB}{0, 110, 184}

\hypersetup{colorlinks,citecolor=nicegreen,linkcolor=nicered,urlcolor=navyblue}

 \newcommand{\clabel}[2][]{#2}
 
\begin{document}


    \title{Constraining the $DDD^*$ three-body bound state via the $Z_c(3900)$ pole}
    \author{Hai-Xiang Zhu\,\orcidlink{0009-0008-7084-7924}}\affiliation{School of Physics, Sun Yat-sen University, Guangzhou 510275, China}
\author{Lu Meng\,\orcidlink{0000-0001-9791-7138}}\email{lu.meng@rub.de}
	\affiliation{Institut f\"ur Theoretische Physik II, Ruhr-Universit\"at Bochum,  D-44780 Bochum, Germany }

	\author{Yao Ma\,\orcidlink{0000-0002-5868-1166}}\email{yaoma@pku.edu.cn}
	\affiliation{School of Physics and Center of High Energy Physics,
		Peking University, Beijing 100871, China}
\author{Ning Li\,\orcidlink{0000-0003-2987-2809}}\email{lining59@mail.sysu.edu.cn}\affiliation{School of Physics, Sun Yat-sen University, Guangzhou 510275, China}
\author{Wei Chen\,\orcidlink{0000-0002-8044-5493}}\email{chenwei29@mail.sysu.edu.cn}
\affiliation{School of Physics, Sun Yat-sen University, Guangzhou 510275, China}
\affiliation{Southern Center for Nuclear-Science Theory (SCNT), Institute of Modern Physics, 
Chinese Academy of Sciences, Huizhou 516000, Guangdong Province, China}

	\author{Shi-Lin Zhu\,\orcidlink{0000-0002-4055-6906}}\email{zhusl@pku.edu.cn}
	\affiliation{School of Physics and Center of High Energy Physics,
		Peking University, Beijing 100871, China}
	
	\begin{abstract}

In this study, we propose using the $Z_c(3900)$ pole position to constrain the existence of the $DDD^*$ three-body bound state within the one-boson-exchange (OBE) model. The existence of the $DDD^*$ bound state remains uncertain due to significant variations in the OBE interaction, particularly in the strength of scalar-meson-exchange interactions, which can differ by a factor about 20 between two commonly used OBE models. This discrepancy renders the $DDD^*$ system highly model-dependent. To address this issue, we constrain the scalar-meson-exchange interaction using the $Z_c(3900)$ pole position, where the pseudoscalar-meson coupling is well-determined, and the $\rho$- and $\omega$-exchange interactions nearly cancel each other out, leaving the coupling constant of the $\sigma$-exchange as the only unknown parameter. Our results indicate that the isospin-$\frac{1}{2}$ $DDD^*$ bound states exist when $Z_c(3900)$ is a virtual state of $D\bar{D}^*/\bar{D}D^*$ located within approximately $-15$ MeV of the threshold. However, the three-body bound state is gone when the $Z_c(3900)$ virtual state pole is more than $20$ MeV away from the threshold. Each experimental progress, either on the $DDD^*$ state or the $Z_c(3900)$, can shed light on the nature of the other state.   Another significant outcome is a refined set of OBE model parameters calibrated using the pole positions of $X(3872)$, $T_{cc}(3875)$, and $Z_c(3900)$, rigorously addressing the cutoff dependence. These parameters provide a valuable resource for more accurate calculations of systems involving few-body $D$, $D^*$ and their antiparticles. Additionally, we find no evidence of the $DDD^*$ three-body resonances after extensive search using a combination of the Gaussian expansion method and the complex scaling method. 
	\end{abstract}
	
	\maketitle
	
	\section{Introduction}~\label{sec:intro}
Over the past two decades, doubly heavy (either the hidden flavor or the open flavor) di-hadron systems have become a hot topic in hadron physics.  This interest stems from the experimental discovery of numerous multiquark candidates since 2003, many of which are frequently interpreted as molecular states formed by two heavy-flavor hadrons, see Refs.~\cite{Chen:2016qju,Esposito:2016noz,Guo:2017jvc,Liu:2019zoy,Brambilla:2019esw,Chen:2022asf,Meng:2022ozq} for reviews. Among these di-hadron states, the $X(3872)$ (also referred to as $\chi_{c1}(3872)$)~\cite{Belle:2003nnu}, $Z_c(3900)$~\cite{BESIII:2013ris,Belle:2013yex}, and $T_{cc}(3875)$~\cite{LHCb:2021vvq,LHCb:2021auc}, stand out as the ``star" examples, which are
considered as two-body systems of $[\DDbar]^{C=+1}$, $[\DDbar]^{C=-1}$ and $\DD$ respectively. Theorists' interests in doubly charmed hadronic molecular states can be traced back to an even earlier time~\cite{Voloshin:1976ap,Tornqvist:1991ks,Tornqvist:1993ng}.

In the di-hadron scenario, the interactions forming doubly heavy molecules are studied using various frameworks, including meson-exchange models~\cite{Tornqvist:1991ks,Tornqvist:1993ng,Gamermann:2006nm,Liu:2009qhy,Liu:2008fh,Liu:2008xz,Ding:2009vj,Sun:2011uh,Thomas:2008ja,Lee:2009hy,Chen:2017vai, Chen:2020yvq, Chen:2021vhg,Dong:2021bvy, Dong:2021juy,Peng:2023lfw,Cheng:2022qcm}, quark models~\cite{Deng:2022cld,Ortega:2022efc,Meng:2023jqk,Ma:2023int,Wu:2024zbx}, QCD sum rules~\cite{Du:2012wp,Wang:2013daa,Wang:2014gwa,Chen:2016jxd}, effective field theories~\cite{Wang:2019nvm,Wang:2019ato,Braaten:2020nmc,Du:2021zzh,Wang:2022jop,Abolnikov:2024key,Chacko:2025hxk}, and lattice QCD~\cite{Chen:2022vpo,Lyu:2023xro,Meng:2023bmz,Meng:2024kkp,Whyte:2024ihh} etc. Notably, the meson-exchange model--often referred to as the one-boson-exchange (OBE) model--has achieved significant success in elucidating heavy-flavor hadronic molecules~\cite{Tornqvist:1991ks,Tornqvist:1993ng,Liu:2009qhy,Liu:2008fh,Liu:2008xz,Ding:2009vj,Sun:2011uh,Thomas:2008ja,Lee:2009hy}, just like its high-precision description of nuclear forces~\cite{Machleidt:1987hj}. This model considers the exchange of mesons such as \(\pi\), \(\eta\), \(\rho\), \(\omega\), and \(\sigma\). Prior to the discovery of \(T_{cc}(3875)\), Li \etal had predicted a very loosely bound \(\DD\) state using the OBE model, employing parameters initially developed in their study of \(X(3872)\)~\cite{Li:2012cs,Li:2012ss}. The interactions governing \(T_{cc}\) and \(X(3872)\) are consistent with G-parity rules. It should be noticed that while doubly heavy tetraquark states have been predicted for over 40 years~\cite{Ader:1981db} (see a concise review in Ref.~\cite{Richard:2022fdc}), explicit predictions of molecular-type doubly charmed tetraquark states were scarce before the landmark LHCb observations~\cite{LHCb:2021vvq,LHCb:2021auc}. Recently, the OBE model was extended to P-wave \(\DD\) and \(\DDbar\) systems, suggesting that the \(G(3900)\) structure, recently observed by BESIII and previously reported by the BaBar~\cite{BaBar:2008drv} and Belle~\cite{Belle:2007qxm} collaborations, corresponds to a P-wave \(\DDbar\) resonance~\cite{Lin:2024qcq}.

Discussions on heavy-flavor di-hadron systems have naturally been extended to three-body systems, particularly given that three-body systems exhibit fascinating features not seen in two-body systems, such as the Efimov effect and the three-body force effect, which remain mysterious even in nucleon systems~\cite{Kalantar-Nayestanaki:2011rzs,Endo:2024cbz}. In Refs.~\cite{Canham:2009zq, Wilbring:2017fwy}, the three-body universality was explored in the charmed and bottom sectors, respectively. The existence of three-body bound states such as \(BBB^*\)~\cite{Ma:2018vhp} and \(DDD^*\)~\cite{Wu:2021kbu} was reported based on interactions modeled using the OBE framework, with the latter prediction inspired by the observation of the \(T_{cc}(3875)\) state. Additionally, investigations have been conducted into systems like \(D^{(*)}D^{(*)}D^{(*)}\)~\cite{Bayar:2022bnc,Pan:2022whr,Ortega:2024ecy,Luo:2021ggs}, \(D^{(*)}D^{(*)}\bar{D}^{(*)}\)~\cite{Tan:2024omp,Valderrama:2018sap}, and \(D^{(*)}B^{(*)}\bar{B}^{(*)}\)~\cite{Dias:2018iuy}, as well as systems where one, two, or three heavy mesons are replaced by kaons~\cite{Ma:2017ery,Ren:2018pcd,Ren:2018qhr,Wu:2019vsy,Zhang:2021hcl,Zhang:2024yfj,Zhai:2024luy,Ren:2024mjh}.  In Ref.~\cite{Contessi:2020jqa}, the prediction of three-body and four-body bound states composed of \(X(3872)\) was discussed. For a comprehensive review of three-body heavy-flavor systems, one can refer to Refs.~\cite{MartinezTorres:2020hus,Liu:2024uxn}.

From a theoretical perspective, there are two complexities which may hinder our ability to make reliable predictions for three-body heavy flavor systems: the uncertain interactions and the computational challenges associated with three-body systems, particularly in the case of resonance states. Even neglecting the complexity from the unknown three-body interactions, the two-body interaction for the heavy hadrons systems can vary significantly across different models. For example, the OBE models for the \(\DD\) and \(\DDbar\) systems presented in Refs.~\cite{Li:2012cs,Li:2012ss} (referred to as model-I) and Refs.~\cite{Liu:2019stu,Wu:2021kbu} (model-II) share very similar analytical expressions. However, the scalar-meson-exchange potential in model-II is nearly 20 times stronger than that in model-I, as illustrated in Table~\ref{tab:potential}. Consequently, the model-II predicts a bound \(DDD^*\) state~\cite{Wu:2021kbu}, whereas calculations using the model-I yield an unbound result (see Sec.~\ref{sec:3-body}). Notably, both interactions provide reasonable and consistent results for the \(X(3872)\) and \(T_{cc}(3875)\). Meanwhile, exactly solving the Faddeev equations~\cite{Faddeev:1960su} for three-body systems is notoriously challenging, particularly for resonance states with finite lifetimes. In such cases, the analyticity of the three-body scattering amplitude becomes highly complex~\cite{Dawid:2023jrj}.

    \begin{table*}[hpt]
    \centering
      \caption{ Momentum space potentials for the \(DD\), \(\DD\), and \(\DDbar\) systems. The superscripts \(D\) and \(C\) denote the direct and cross diagrams, respectively. Two sets of \(C_\text{coupling}\) values are provided: model-I from Refs.~\cite{Li:2012cs,Li:2012ss}, used as the baseline for this work, and model-II from Refs.~\cite{Liu:2019stu,Wu:2021kbu}. The terms \(\bm{\epsilon}'^* \) and \(\bm{\epsilon}\) represent the polarization vectors of the final and initial vector mesons, respectively. For \(DD\) systems, the substitution \(\bm{\epsilon}'^* \cdot \bm{\epsilon} \to 1\) is required. In the direct diagram, \(\bm{q} = \bm{p}' - \bm{p}\), while in the cross diagram, \(\bm{k} = \bm{p}' + \bm{p}\), where \(\bm{p}'\) and \(\bm{p}\) are the momenta of the final and initial states, respectively. Here, \(u\) denotes the effective mass of the exchanged meson. Two additional isospin operators are defined as \(\bm{\tau}_{1} \cdot \bm{\tau}_{2}^{C} \equiv (3I - \bm{\tau}_{1} \cdot \bm{\tau}_{2})/2\) and \(I^{C} \equiv (I + \bm{\tau}_{1} \cdot \bm{\tau}_{2})/2\). The potential in coordinate space can be obtained through Fourier transformation, as detailed in Appendix~\ref{app:fourier}. }
    \label{tab:potential}  
 \begin{tabular*}{\hsize}{@{}@{\extracolsep{\fill}}cccccccccc@{}}
\hline 
\hline 
\multirow{2}{*}{$V$}& \multicolumn{4}{c}{$C_{\text{coupling}}$} & \multirow{2}{*}{$\mathcal{O}_{r,s}$} & \multicolumn{4}{c}{$\mathcal{O}_{iso}$}\tabularnewline
 & \multicolumn{2}{c}{Model-I\cite{Li:2012ss,Li:2012cs}} & \multicolumn{2}{c}{Model-II\cite{Liu:2019stu,Wu:2021kbu}} &  & $DD^{*}$ & $DD$ & $[\DDbar]^{C=+1}$ & $[\DDbar]^{C=-1}$\tabularnewline
\hline 
$V_{\rho}^{D}$ & $\frac{\beta^{2}g_{v}^{2}}{2}$ & 13.62 & $2g_{\rho}^{2}$ & 13.52 & \multirow{3}{*}{$-\frac{\bm{\epsilon}'^{*}\cdot\bm{\epsilon}}{u^{2}+\bm{q}^{2}}$} & $-\frac{\bm{\tau}_{1}\cdot\boldsymbol{\tau}_{2}}{2}$ & $-\frac{\boldsymbol{\tau}_{1}\cdot\bm{\tau}_{2}}{2}$ & $-\frac{\bm{\tau}_{1}\cdot\bm{\tau}_{2}}{2}$ & $-\frac{\bm{\tau}_{1}\cdot\bm{\tau}_{2}}{2}$\tabularnewline
$V_{\omega}^{D}$ & $\frac{\beta^{2}g_{v}^{2}}{2}$ & 13.62 & $2g_{\omega}^{2}$ & 13.52 &  & $-\frac{1}{2}I$ & $-\frac{1}{2}I$ & $\frac{1}{2}I$ & $\frac{1}{2}I$\tabularnewline
$V_{\sigma}^{D}$ & $g_{s}^{2}$ & 0.58 & $g_{\sigma}^{2}$ & 11.56 &  & $I$ & $I$ & $I$ & $I$\tabularnewline
\hline 
$V_{\pi}^{C}$ & $\frac{g_a^{2}}{f_{\pi}^{2}}$ & 19.15 & $\frac{g^{2}}{f_{\pi}^{2}}$ & 20.66 & \multirow{2}{*}{$\frac{\left(\bm{k}\cdot\bm{\epsilon}'^{*}\right)\left(\bm{k}\cdot\epsilon\right)}{u^{2}+\bm{k}^{2}}$} & $-\frac{\bm{\tau}_{1}\cdot\bm{\tau}_{2}^{C}}{2}$ & 0 & $\frac{\bm{\tau}_{1}\cdot\bm{\tau}_{2}}{2}$ & $-\frac{\bm{\tau}_{1}\cdot\bm{\tau}_{2}}{2}$\tabularnewline
$V_{\eta}^{C}$ & $\frac{g_a^{2}}{f_{\pi}^{2}}$ & 19.15 & 0 & 0 &  & $-\frac{1}{6}I^{C}$ & 0 & $-\frac{1}{6}I$ & $\frac{1}{6}I$\tabularnewline
$V_{\rho}^{C}$ & $2\lambda^{2}g_{v}^{2}$ & 21.10 & $\frac{f_{\rho}^{2}}{2M^{2}}$ & 19.64 & \multirow{2}{*}{$\frac{\left(\bm{k}\cdot\bm{\epsilon}'^{*}\right)\left(\bm{k}\cdot\epsilon\right)-\bm{k}^{2}\left(\bm{\epsilon}'^{*}\cdot\bm{\epsilon}\right)}{u^{2}+\bm{k}^{2}}$} & $\frac{\bm{\tau}_{1}\cdot\bm{\tau}_{2}^{C}}{2}$ & 0 & $\frac{\bm{\tau}_{1}\cdot\bm{\tau}_{2}}{2}$ & $-\frac{\bm{\tau}_{1}\cdot\bm{\tau}_{2}}{2}$\tabularnewline
$V_{\omega}^{C}$ & $2\lambda^{2}g_{v}^{2}$ & 21.10 & $\frac{f_{\omega}^{2}}{2M^{2}}$ & 19.64 &  & $\frac{1}{2}I^{C}$ & 0 & $-\frac{1}{2}I$ & $\frac{1}{2}I$\tabularnewline
\hline 
\hline 
\end{tabular*}
\end{table*}

        In this work, we focus on the $DDD^*$ three-body system. While the existence of three-body bound states in this system has been explored in Ref.~\cite{Wu:2021kbu}, it has not yet been definitively established, especially considering possible variations in the OBE interaction, particularly the scalar-exchange contributions. To address this, we constrain the scalar-meson-exchange interactions using the pole position of the $Z_c(3900)$. For $Z_c(3900)$, the pseudoscalar-meson coupling is well-determined and the $\rho$- and $\omega$-exchange interactions are expected to largely cancel each other, leaving the coupling constant of the $\sigma$ exchange as the unique unknown parameter. Additionally, the coupling constants in OBE models are inherently cutoff-dependent, a factor often overlooked in previous literature, leading to ambiguity in predictions. In this work, we address this issue by adopting definite cutoffs and recalibrating the coupling constants for vector meson and scalar meson exchanges using the pole positions of $X(3872)$, $T_{cc}(3875)$, and $Z_c(3900)$. Coupling constants corresponding to cutoff values ranging from 1.00 to 1.35 GeV, in increments of 0.05 GeV, are provided, which serve as a valuable resource for more precise calculations in few-body systems involving $D$, $D^*$ and their antiparticles. Meanwhile, the combination of the Gaussian expansion method (GEM)~\cite{Hiyama:2003cu} and the complex scaling method (CSM)~\cite{Aguilar:1971ve,Balslev:1971vb,Moiseyev:1998gjp,Aoyama:2006hrz,Carbonell:2013ywa,Hiyama:2016nwn,Dote:2017wkk,Lin:2022wmj,Happ:2023kcc,Chen:2023eri,Meng:2024yhu,Ma:2024vsi,Wu:2024euj,Wu:2024zbx} has proven to be a powerful tool for studying few-body resonance systems. In this work, we also investigate the potential existence of $DDD^*$ resonances using CSM within Gaussian bases.

    The paper is organized as follows. In Sec.~\ref{sec:obe}, the OBE interaction is introduced, and the model parameters are determined. In Sec.~\ref{sec:3-body}  methodology for studying $DDD^*$ three-body systems along with the numerical results is presented. The final conclusions are provided in Sec.~\ref{sec:concl}. Appendix~\ref{app:fourier} details the Fourier transformation used to derive the potential in coordinate space from its momentum-space representation. Appendix B offers supplementary information on the spatial structures of the \(DDD^*\) and \(T_{cc}(3875)\) systems.

\section{One-boson-exchange interaction}~\label{sec:obe}
\subsection{Analytical results}

Using the heavy quark spin symmetry, the pseudoscalar  meson $D$ and vector meson $D^*$ can be formulated in a superfield $\mathcal{H}$~\cite{Georgi:1990cx,Mannel:1990vg,Falk:1991nq,Liu:2008fh,Liu:2009qhy,Li:2012cs,Li:2012ss},
\begin{equation}
        {\cal H}=\frac{1+\slashed{v}}{2}(P_{\mu}^*\gamma^{\mu}-P\gamma_{5}),
\end{equation}
where $P=(D^0,D^+)$ and $P_\mu^*=(D^{*0},D^{*+})_\mu$. The velocity of the heavy meson is denoted by $v=(1,0,0,0)$. Similarly, their antiparticles can be described by the superfield $\tilde{{\cal H}}$, defined as,
\begin{eqnarray}
  \tilde{{\cal H}}=(\tilde{P}_{\mu}^*\gamma^{\mu}-\tilde{P}\gamma_{5})\frac{1-\slashed{v}}{2},
\end{eqnarray}
with $\tilde{P}=(\bar{D}^0,D^-)^\text{T}$ and $\tilde{P}_\mu^*=(\bar{D}^{*0},D^{*-})_\mu^\text{T}$. Here, we choose the convention of charge conjugation, $D\xrightarrow{\mathcal{C}}\bar{D}$ and $D^*\xrightarrow{\mathcal{C}}-\bar{D}^*$, namely $\mathcal{H}\xrightarrow{\mathcal{C}}C^{-1}\tilde{\mathcal{H}}^\text{T}C$, where $C=i\gamma^2\gamma^0$ is the charged conjugation matrix. 
 The conjugations of $\cal{H}$ and $\tilde{\mathcal{H}}$ are defined as $\bar{\mathcal{H}}=\gamma_0\mathcal{H}^\dagger\gamma_0$ and $\bar{\tilde{\mathcal{H}}}=\gamma_0\tilde{\mathcal{H}}^\dagger\gamma_0$.

The Lagrangians in the OBE model read,
\begin{eqnarray}
    &\mathcal{L}&=g_s\Tr\left[\mathcal{H}\sigma\bar{\mathcal{H}}\right]+ig_a\Tr\left[\mathcal{H}\gamma_\mu\gamma_5\mathcal{A}^\mu\bar{\mathcal{H}}\right]\nonumber\\
    &&+i\beta\Tr\left[\mathcal{H} v_\mu (\mathcal{V}^\mu-\rho^\mu)\bar{\mathcal{H}}\right]+i\lambda\Tr\left[\mathcal{H}\sigma_{\mu\nu}F^{\mu\nu}\bar{\mathcal{H}}\right]\nonumber\\
    &&+g_s\Tr\left[\bar{\tilde{\mathcal{H}}}\sigma\tilde{\mathcal{H}}\right]+ig_a\Tr\left[\bar{\tilde{\mathcal{H}}}\gamma_\mu\gamma_5\mathcal{A}^\mu\tilde{\mathcal{H}}\right]\nonumber\\
    &&-i\beta\Tr\left[\bar{\tilde{\mathcal{H}}} v_\mu (\mathcal{V}^\mu-\rho^\mu)\tilde{\mathcal{H}}\right]+i\lambda\Tr\left[\bar{\tilde{\mathcal{H}}}\sigma_{\mu\nu}F^{\mu\nu}\tilde{\mathcal{H}}\right].
\end{eqnarray}
where $F^{\mu\nu}=\partial^\mu\rho^\nu-\partial^\nu\rho^\mu-[\rho^\mu,\rho^\nu]$ represents the field strength tensor of vector mesons,  
$\mathcal{V}^\mu$ and $\mathcal{A}^\mu$ are the vector and axial building blocks of pseudoscalar mesons respectively,
\begin{eqnarray}
    &&\mathcal{V}^\mu=\frac{1}{2}[\xi^\dagger,\partial_\mu\xi],\; \mathcal{A}^\mu=\frac{1}{2}\{\xi^\dagger,\partial_\mu\xi\},\;
    \xi=\exp(i\mathbb{P}/f_\pi).
\end{eqnarray}
 The pseudoscalar meson matrix $\mathbb{P}$ is defined as
\begin{eqnarray}
\mathbb{P}=
\begin{bmatrix}\frac{\pi^{0}}{\sqrt{2}}+\frac{\eta}{\sqrt{6}} & \pi^{+}\\
\pi^{-} & -\frac{\pi^{0}}{\sqrt{2}}+\frac{\eta}{\sqrt{6}}
\end{bmatrix}.\nonumber\\
\end{eqnarray}
The multiplet of the vector meson fields $\rho^\mu$ is
\begin{eqnarray}
     \rho^\mu=\frac{ig_V}{\sqrt{2}}\begin{bmatrix}\frac{\rho^{0}+\omega}{\sqrt{2}}
 & \rho^{+}\\
\rho^{-} & \frac{-\rho^{0}+\omega}{\sqrt{2}}
\end{bmatrix}^\mu.
\end{eqnarray}
In principle, the iso-triplet \(\rho\) (\(\pi\)) and the iso-singlet \(\omega\) (\(\eta\)) belong to different multiplets under the flavor SU(2) symmetry. However, to reduce the number of coupling constants, we group them into the same matrix, taking into account their relationship within the flavor SU(3) symmetry. 

\begin{table}[]
    \centering
        \caption{Hadron masses and coupling constants in model-I~\cite{Li:2012cs,Li:2012ss,ParticleDataGroup:2024cfk}. All masses and $f_\pi$ are given in units of GeV, while $\lambda$ is expressed in $\text{GeV}^{-1}$. The present values of $g_a$ and $f_\pi$ have been updated based on recent experimental results, leading to slight differences compared to those in Refs.~\cite{Li:2012cs,Li:2012ss}. The values of $g_s$, $\lambda$, and $\beta$ are used as baseline parameters, where a rescaling factor can be applied, as specified in Eq.~\eqref{eq:coupling-ratio}.  }
    \label{tab:lag_para}
\begin{tabular*}{\hsize}{@{}@{\extracolsep{\fill}}cccccccc@{}}
\hline 
\hline 
\multirow{2}{*}{Mass } & $m_{D}$ & $m_{D^{*}}$ & $m_{\pi}$ & $m_{\eta}$ & $m_{\rho}$ & $m_{\omega}$ & $m_{\sigma}$\tabularnewline
\cline{2-8} 
 & 1.867 & 2.009 & 0.137 & 0.548 & 0.775 & 0.783 & 0.600\tabularnewline
\hline 
\multirow{2}{*}{Coupling} & $f_{\pi}$ & $g_{a}$ & $g_{V}$ & $\beta$ & $\lambda$  & $g_{s}$ & \tabularnewline
\cline{2-8} 
 & 0.13025 & 0.57 & 5.8 & 0.9 & 0.56  & 0.76 & \tabularnewline
\hline 
\hline 
\end{tabular*}
\end{table}

In the pseudoscalar sector of the Lagrangians, the parameters are well-determined. The pion decay constant \( f_\pi \) is a well-known quantity, and the axial coupling constant \( g_a \) can be accurately extracted from the \( D^* \) width. However, the remaining parameters can only be determined through model-dependent methods. As an example, we adopt the values from Refs.~\cite{Li:2012cs,Li:2012ss} (model-I), which are summarized in Table~\ref{tab:lag_para}. In the vector sector, there are three parameters: \( g_V \), \( \beta \), \( \lambda \), and an additional scalar coupling \( g_s \). However, only three of them are independent, as \( g_V \) acts as an overall factor in the vector sector. In model-I~\cite{Li:2012cs,Li:2012ss}, these couplings are determined using vector meson dominance, combined with results from lattice QCD and light-cone sum rules~\cite{Isola:2003fh}. In the scalar sector, the \(\sigma\) mass and couplings are derived from the \(\Sigma\)-model~\cite{Bardeen:2003kt,Liu:2008xz}. Notably, the determination of these parameters is not achieved within an unified framework. In other studies, such as Refs.~\cite{Ding:2009vj,Kim:2019rud,Liu:2019stu}, similar Lagrangians have been constructed, but the coupling constants were determined using different approaches, leading to variations in numerical values. In this work, we adopt the values of model-I listed in Table~\ref{tab:lag_para} as a baseline. To explore the impact of varying coupling constants in the scalar and vector sectors, we introduce three ratio factors,
\begin{equation}
    \lambda\to\lambda R_{\lambda},\quad\beta\to\beta R_{\beta},\quad g_{s}\to g_{s}R_{s}.\label{eq:coupling-ratio}
\end{equation}

\begin{figure}
    \centering
    \includegraphics[width=0.45\textwidth]{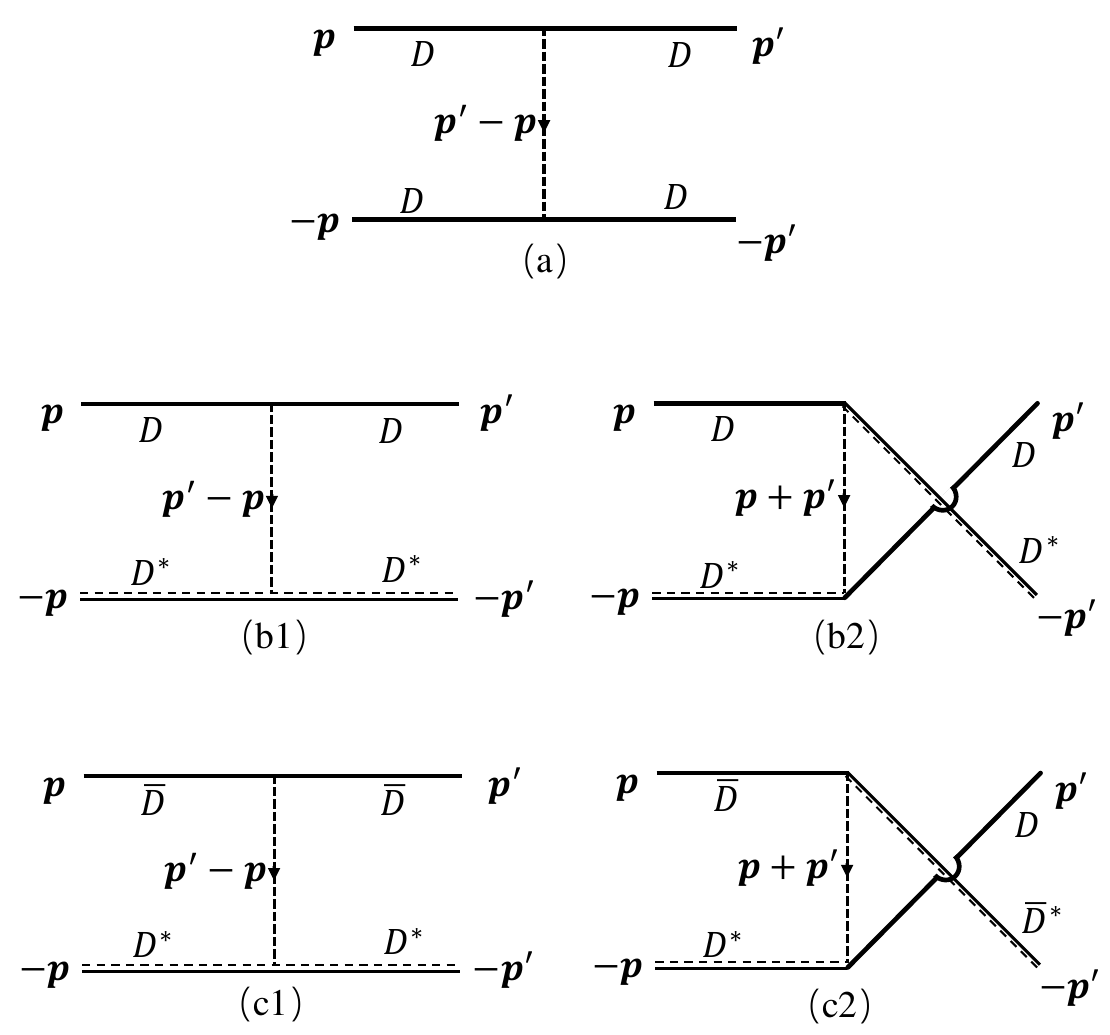}
    \caption{Feynman diagrams to derive OBE potentials: (a) for $DD$, (b1) and (b2) for $\DD$, and (c1) and (c2) for $\DDbar$. Diagrams (b1) and (c1) are referred to as direct diagrams, while (b2) and (c2) are referred to as cross diagrams.}
    \label{fig:feynman}
\end{figure}

The OBE potentials can be derived from two-body scattering Feynman diagrams, as shown in Fig.~\ref{fig:feynman}: (a) for $DD$, (b1) and (b2) for $\DD$, and (c1) and (c2) for $\DDbar$. Diagrams (b1) and (c1) are referred to as direct diagrams, while (b2) and (c2) are referred to as cross diagrams. The analytical expressions for the potentials in momentum space can be represented as  
\begin{equation}    
V = C_{\text{coupling}} \times {\cal O}_{r,s} \times \mathcal{O}_{\text{iso}},
\end{equation}  
where the ingredients are detailed in Table~\ref{tab:potential}. It is important to note that the exchanged momenta differ between the direct and cross diagrams: for the direct diagram, $\bm{q} = \bm{p}' - \bm{p}$, while for the cross diagram, $\bm{k} = \bm{p}' + \bm{p}$. This distinction affects systems with P-wave (and other odd partial waves), as discussed in the Supplemental Materials of Ref.~\cite{Lin:2024qcq} and Appendix~\ref{app:fourier}. In Table~\ref{tab:potential}, $u$ represents the effective mass of the exchanged meson, derived as follows: 
\begin{equation}
-\frac{1}{k_{0}^{2} - \bm{k}^{2} - m^{2}} = \frac{1}{\bm{k}^{2} + (m^{2} - k_{0}^{2})} = \frac{1}{\bm{k}^{2} + u^{2}},
\end{equation}  
using the cross diagram as an example. In the static limit, $k_0$ and $q_0$ are determined by the mass differences between the initial and final states. Specifically, $q_0^2 = 0$ and $k_0^2 = (m_{D^*} - m_D)^2$. Notably, $k_0^2$ is very close to $m_{\pi}^2$~\footnote{In channels with specific charges, $k_0^2 > m_{\pi}^2$ indicates the opening of $D\bar{D}\pi$ three-body threshold. Such effects are neglected in this work.} and much smaller than other meson masses. Thus, we set $u = 0$ for one-pion-exchange (OPE) interaction and $u = m$ for other cases. 
The isospin-averaged particle masses are taken from the Review of Particle Physics~\cite{ParticleDataGroup:2024cfk} and are listed in Table~\ref{tab:lag_para}. In principle, the masses of the exchanged mesons in OBE do not correspond exactly to their pole masses. This is particularly true for vector and scalar mesons, which are resonances with finite widths, however, only real masses are adopted in OBE interactions. These masses should be viewed as parameters that model medium- and short-range interactions. Vector and scalar meson masses and their couplings can be derived by matching the OBE amplitude to the correlated two-pion contribution in effective field theories at low energy~\cite{Kim:2019rud,Wu:2023uva}. However, in this work, we fix the meson masses to reduce the number of parameters.

\begin{figure*}
    \centering
    \includegraphics[width=0.85\textwidth]{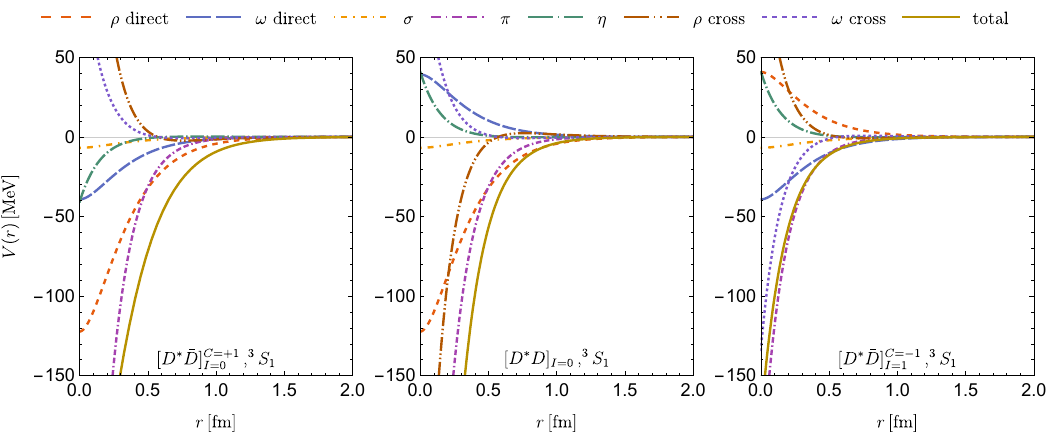}
    \caption{The coordinate space potentials  in the isospin limit for the systems corresponding to the $X(3872)$, $T_{cc}(3875)$, and $Z_c(3900)$ states. The parameter $\Lambda$ is set to 1.20 GeV, with $R_{\beta}$, $R_{\lambda}$, and $R_s$ fixed at 1.}
    \label{fig:V-baseline}
\end{figure*}

The potential in coordinate space can be readily derived from its counterpart in momentum space; see Appendix~\ref{app:fourier} for details. Potentials obtained directly from Feynman diagrams are often singular, featuring terms like the $\delta$ function. To regulate these singularities, this work introduces a regulator to suppress contributions from the high-momentum region:
\begin{eqnarray}
    V^{D}(\bm{q},u)&\to&V^{D}(\bm{q},u)F^{2}(u,\Lambda,\bm{q}^{2}),\nonumber\\
    V^{C}(\bm{k},u)&\to&V^{C}(\bm{k},u)F^{2}(u,\Lambda,\bm{k}^{2}),\nonumber\\
    F(u,\Lambda,\bm{q}^{2})&=&\frac{\Lambda^{2}-u^{2}}{\Lambda^{2}+\bm{q}^{2}}. \label{eq:regulator}
\end{eqnarray}
Here, $\Lambda$ acts as a new parameter, introducing additional model dependence.

In Table~\ref{tab:iso_factor}, the specific values of isospin factors are presented. Given the significant isospin violation observed in the decay patterns of $X(3872)$~\cite{Belle:2005lfc,BaBar:2010wfc,BESIII:2019qvy}, the $\DD$ and $\DDbar$ systems can be studied under two scenarios: the exact isospin symmetry limit one and the isospin symmetry violating one. In the isospin-violating scenario, the mass differences between channels with different charges are taken into account, rendering isospin no longer a quantum number of the energy eigenstates. Conversely, in the isospin symmetry limit, these mass splittings are neglected. For example, the relevant mass splittings for the $X(3872)$ and $T_{cc}(3875)$ systems are:  
\begin{eqnarray}
\delta_{X}&=&m_{D^{+}}+m_{D^{*+}}-m_{D^{0}}-m_{D^{*0}}=8.2\text{ MeV},\\
\delta_{T}&=&m_{D^{+}}+m_{D^{*0}}-m_{D^{0}}-m_{D^{*+}}=1.4\text{ MeV}.
\end{eqnarray}
In Fig.~\ref{fig:V-baseline}, the coordinate space potentials in the isospin symmetry limit for the systems corresponding to the $X(3872)$, $T_{cc}(3875)$, and $Z_c(3900)$ states are presented. One can clearly observe the contributions to the potential from different meson exchanges. In Table~\ref{tab:iso_factor}, both the isospin factors in the isospin symmetry limit scenario and the isospin-violating scenarios are given. Notably, for the isovector $[\DDbar]^{C=-1}_{I=1}$ system, which corresponds to the $Z_{c}(3900)$ state, the isospin factors for $\rho$ and $\omega$ exchanges are equal in magnitude but opposite in sign for both cross and direct diagrams. As a result, if the mass difference between \(\rho\) and \(\omega\) is neglected, the vector-meson exchange vanishes entirely. In our calculations, the specific mass values of $\rho$ and $\omega$ in the model are listed in Table~\ref{tab:lag_para} as $0.775$ GeV and $0.783$ GeV, respectively, which are nearly identical. Numerically, in the third sub-figure of Fig.~\ref{fig:V-baseline} (the exchange interaction potential of $Z_c(3900)$), we can observe that the exchange interaction potentials of the $\rho$ meson and $\omega$ meson are nearly equal in magnitude and opposite in sign. Consequently, the structure of \(Z_{c}(3900)\) becomes sensitive to the \(\sigma\)-exchange interactions. The \(\sigma\)-exchange interaction is isospin-independent, as shown in Table~\ref{tab:iso_factor}, and provides an attractive force (as shown in Fig.~\ref{fig:V-baseline}) for all types of particle pairs. This suggests that the strong \(\sigma\)-interaction facilitates the formation of the few-body bound states.

In this work, three-body force is neglected. Within the framework of chiral effective field theory, the three-body nuclear force first emerges at the next-to-next-to-leading order according to the Weinberg power counting scheme~\cite{Epelbaum:2009sd,Machleidt:2011zz}, and thus its contribution is suppressed compared to the two-body interaction. Furthermore, {\rm ab initio} calculations for light nuclei have demonstrated that the three-body force typically increases the binding energy by approximately 10–25\%~\cite{Pieper:2004qh}. Given this, we anticipate that the impact of the three-body force is unlikely to dominate over other sources of uncertainty in our analysis. Therefore, we have chosen to neglect its contribution.

\begin{table*}[]
    \centering
        \caption{Isospin factor for $\DDbar$ and $\DD$ systems. The ``diag." and ``off." are the diagonal and off-diagonal  matrix element in the isospin violating scenario. }
    \label{tab:iso_factor}
\begin{tabular*}{\hsize}{@{}@{\extracolsep{\fill}}ccccccccccccc@{}}
\hline 
\hline 
 & \multicolumn{4}{c}{$DD^{*}$} & \multicolumn{4}{c}{$\DDbar,C=+1$} & \multicolumn{4}{c}{$\DDbar,C=-1$}\tabularnewline
$\langle\mathcal{O}_{iso}\rangle$ & diag. & off. & $I=0$ & $I=1$ & diag. & off. & $I=0$ & $I=1$ & diag. & off. & $I=0$ & $I=1$\tabularnewline
\hline 
$V_{\rho}^{D}$ & $\frac{1}{2}$ & $-1$ & $\frac{3}{2}$ & $-\frac{1}{2}$ & $\frac{1}{2}$ & $1$ & $\frac{3}{2}$ & $-\frac{1}{2}$ & $\frac{1}{2}$ & $1$ & $\frac{3}{2}$ & $-\frac{1}{2}$\tabularnewline
$V_{\omega}^{D}$ & $-\frac{1}{2}$ & $0$ & $-\frac{1}{2}$ & $-\frac{1}{2}$ & $\frac{1}{2}$ & $0$ & $\frac{1}{2}$ & $\frac{1}{2}$ & $\frac{1}{2}$ & $0$ & $\frac{1}{2}$ & $\frac{1}{2}$\tabularnewline
$V_{\sigma}^{D}$ & $1$ & $0$ & $1$ & $1$ & $1$ & $0$ & $1$ & $1$ & $1$ & $0$ & $1$ & $1$\tabularnewline
\hline 
$V_{\pi}^{C}$ & $-1$ & $\frac{1}{2}$ & $-\frac{3}{2}$ & $-\frac{1}{2}$ & $-\frac{1}{2}$ & $-1$ & $-\frac{3}{2}$ & $\frac{1}{2}$ & $\frac{1}{2}$ & $1$ & $\frac{3}{2}$ & $-\frac{1}{2}$\tabularnewline
$V_{\eta}^{C}$ & $0$ & $-\frac{1}{6}$ & $\frac{1}{6}$ & $-\frac{1}{6}$ & $-\frac{1}{6}$ & $0$ & $-\frac{1}{6}$ & $-\frac{1}{6}$ & $\frac{1}{6}$ & $0$ & $\frac{1}{6}$ & $\frac{1}{6}$\tabularnewline
$V_{\rho}^{C}$ & $1$ & $-\frac{1}{2}$ & $\frac{3}{2}$ & $\frac{1}{2}$ & $-\frac{1}{2}$ & $-1$ & $-\frac{3}{2}$ & $\frac{1}{2}$ & $\frac{1}{2}$ & $1$ & $\frac{3}{2}$ & $-\frac{1}{2}$\tabularnewline
$V_{\omega}^{C}$ & $0$ & $\frac{1}{2}$ & $-\frac{1}{2}$ & $\frac{1}{2}$ & $-\frac{1}{2}$ & $0$ & $-\frac{1}{2}$ & $-\frac{1}{2}$ & $\frac{1}{2}$ & $0$ & $\frac{1}{2}$ & $\frac{1}{2}$\tabularnewline
\hline 
\hline 
\end{tabular*}
\end{table*}

\subsection{Fix the parameters}
Even though one set of parameters for the OBE potential (model-I) is provided in Table~\ref{tab:lag_para}, the model dependence and uncertainties associated with these parameters have not been seriously evaluated. Notably, the coupling constants reported in the literature exhibit significant variation~\cite{Li:2012cs,Li:2012ss,Ding:2009vj,Kim:2019rud,Liu:2019stu}. As shown in Table~\ref{tab:potential}, the strength of the $\sigma$-exchange potential in model-II \cite{Liu:2019stu} are almost 20 times larger than that in model-I~\cite{Li:2012cs,Li:2012ss}.  Consequently, the  interaction of model-II predicts a bound \(DDD^*\) state~\cite{Wu:2021kbu}, whereas calculations using the interaction of model-I yield an unbound result (see Sec.~\ref{sec:3-body}). Additionally, uncertainties arising from cutoff dependence remain largely unaddressed. A compromise approach is to fix the coupling constants and vary the cutoff within a reasonably assumed range to examine the changes in predictions. This leads to the fact that many theoretical predictions based on the OBE model are highly sensitive to the cutoff. In fact, during the determination of nuclear forces using the OBE model, the coupling constants themselves are cutoff-dependent. Only when the cutoff associated with the coupling constants is explicitly specified does the OBE model possess genuine predictive power.

In this work, we propose using the pole positions of $X(3872)$, $T_{cc}(3875)$, and $Z_c(3900)$ to determine the three independent coupling constants in the OBE models—equivalently, the parameters $R_\lambda$, $R_\beta$, and $R_s$ in Eq.~\eqref{eq:coupling-ratio}. 
Recent refined analyses of the line shapes by LHCb~\cite{LHCb:2020xds} and BESIII~\cite{BESIII:2023hml} suggest that the pole of $X(3872)$ is located on the physical sheet of the $D^0\bar{D}^{*0}/\bar{D}^0D^{*0}$ channel. This implies that $X(3872)$ is likely a loosely bound state of $\DDbar$, assuming the effects of other coupled channels such as $D\bar{D}\pi$, $D\bar{D}\gamma$, $J/\psi \pi\pi$, and $J/\psi\pi\pi\pi$ are negligible. The decay fractions of the final states \( J/\psi \pi\pi \) and \( J/\psi \pi\pi\pi \) are both less than 5\%, making it reasonable to assume that the coupled-channel effects from these hidden charm channels are negligible. Additionally, while a significant decay fraction for the \( D^0\bar{D}^0\pi \) channel has been reported, it is predominantly mediated via \( D^0\bar{D}^{*0} \), as demonstrated in Refs.~\cite{Voloshin:2003nt, Fleming:2007rp}, where the pionic dynamics are expected to play a subleading role. For $T_{cc}(3875)$, the experimental pole lies approximately 0.4 MeV below the $D^0D^{*+}$ threshold~\cite{LHCb:2021auc,LHCb:2021vvq}, strongly indicating that it is a bound state candidate of $\DD$. \clabel[3bodyeffect]{
It has been demonstrated that considering the coupled-channel effects from \(DD\pi\) and \(DD\gamma\) can significantly impact the width of the \(T_{cc}\) state~\cite{Zhang:2024dth,Du:2021zzh}. In fact, in earlier work, \(DD\pi\) and \(DD\gamma\)  introduced as a perturbation in Ref.~\cite{Meng:2021jnw}, already provide an accurate description of the width of \( T_{cc}(3875) \). Therefore, the three-body effects are expected to be subleading and are neglected in this work. In the future, to achieve a more precise determination of the width of the \(DDD^*\) system, the four-body \(DDD\pi\) and \(DDD\gamma\) channels should also be incorporated properly. } Compared to $X(3872)$ and $T_{cc}(3875)$, the pole properties of $Z_c(3900)$ are less definite. However, several recent studies suggest that the pole of $Z_c(3900)$ corresponds to a virtual state~\cite{Albaladejo:2015lob,Nakamura:2023obk,Yu:2024sqv}. This virtual state is located on the unphysical Riemann sheet and lies several tens of MeV below the $\DDbar$ threshold, though alternative theoretical interpretations remain~\cite{Pilloni:2016obd,Chen:2023def}.  It should be noticed that, the threshold cusp interpretation is not in conflict with the virtual state picture. The effect of a virtual state below the threshold is to generate an enhancement at the threshold. It has been demonstrated that the cusp effect could produce a significant bump in the experimental line shape, which typically requires the amplifying influence of a nearby pole (bound state or virtual state)~\cite{Guo:2019twa}. As for other interpretations, such as the triangle singularity or the compact tetraquark scenario, there is no direct connection between the $Z_c(3900)$ and the $DDD^*$ pole. However, a potential significance of this work lies in establishing a relationship between the $Z_c(3900)$ and the $DDD^*$ state exclusively within the virtual state interpretation (or cusp effect). Investigating this relationship could further aid in unraveling the structure of the $Z_c(3900)$. 

Building on the above progress, we treat $X(3872)$ and $T_{cc}(3875)$ as bound states of $\DDbar$ and $\DD$, respectively, and $Z_c(3900)$ as a virtual state of $\DDbar$. Specifically, we adopt the following pole positions relative to their respective thresholds:
\begin{eqnarray}
X(3872)&:&\;(-1)\times(0.2,0.4,0.6)^{B}\text{ MeV},\nonumber\\
T_{cc}(3875)&:&\;(-1)\times(0.2,0.4,0.6)^{B}\text{ MeV},\label{eq:pole-position}
\\Z_{c}(3900)&:&\;(-1)\times(5,10,15,20,25,30,35)^{V}\text{ MeV},\nonumber
\end{eqnarray}
here the superscripts “$B$” and “$V$” denote bound states and virtual states, respectively. Based on these positions, we generate a total of $3 \times 3 \times 7 = 63$ different combinations, allowing us to numerically determine 63 sets of $R_\lambda$, $R_\beta$, and $R_s$. Notably, the binding energies of $X(3872)$ and $T_{cc}(3875)$ are set to be below 1 MeV. This is valid only in the isospin-violating scenario, as the isospin mass splitting exceeds the binding energies. Thus, for $X(3872)$ and $T_{cc}(3875)$, a two-channel formalism that incorporates the isospin mass splitting is employed. The specified pole positions are defined relative to the lower thresholds. For $Z_c(3900)$, whose pole varies across a wide range, the isospin-violating effect is neglected due to the larger uncertainties in its pole position. Therefore, $Z_c(3900)$ is evaluated within the isospin-limit scenario. While studies such as Refs.~\cite{Albaladejo:2015lob,Nakamura:2023obk,Yu:2024sqv} suggest that the $Z_c(3900)$ pole may locate lower than -35 MeV relative to the threshold, we consider such poles unlikely to have observable significance in the line shape. Therefore, we restrict the pole to no more than 35 MeV below the threshold.

In the fitting procedure, we solve the two-body Schrödinger equation in momentum space using deformed contours~\cite{Hagen:2003ev,Hagen:2006dt,Chen:2023eri}:
\[
\frac{p^{2}}{2\mu}\psi(p) + \int_{C}\frac{dq}{(2\pi)^{3}}q^{2}V_{l=0}(p,q)\psi(q) = E\psi(p),
\]
where \(V_{l=0}\) is the S-wave potential following the same convention as described in Ref.~\cite{Lin:2024qcq}, and \(C\) represents the different contours in momentum space. As demonstrated in Refs.~\cite{Hagen:2003ev,Hagen:2006dt,Chen:2023eri}, various contours allow for the extraction of bound state, resonance, and virtual state poles. In this work, we use a straight-line contour, where \(p\) ranges from 0 to infinity, to locate the bound state poles. To extract the virtual state pole, we adopt the contour defined in Ref.~\cite{Chen:2023eri}. During the calculation, the small imaginary part of the pole arising from the left-hand cut is neglected. To solve the Schrödinger equation in momentum space, the contours are discretized, transforming the equation into an eigenvalue problem. We solve these three systems separately, adjusting the coupling constants until the calculated poles match the target values, thus completing the fitting procedure.

\begin{figure*}
    \centering
    \includegraphics[width=0.90\textwidth]{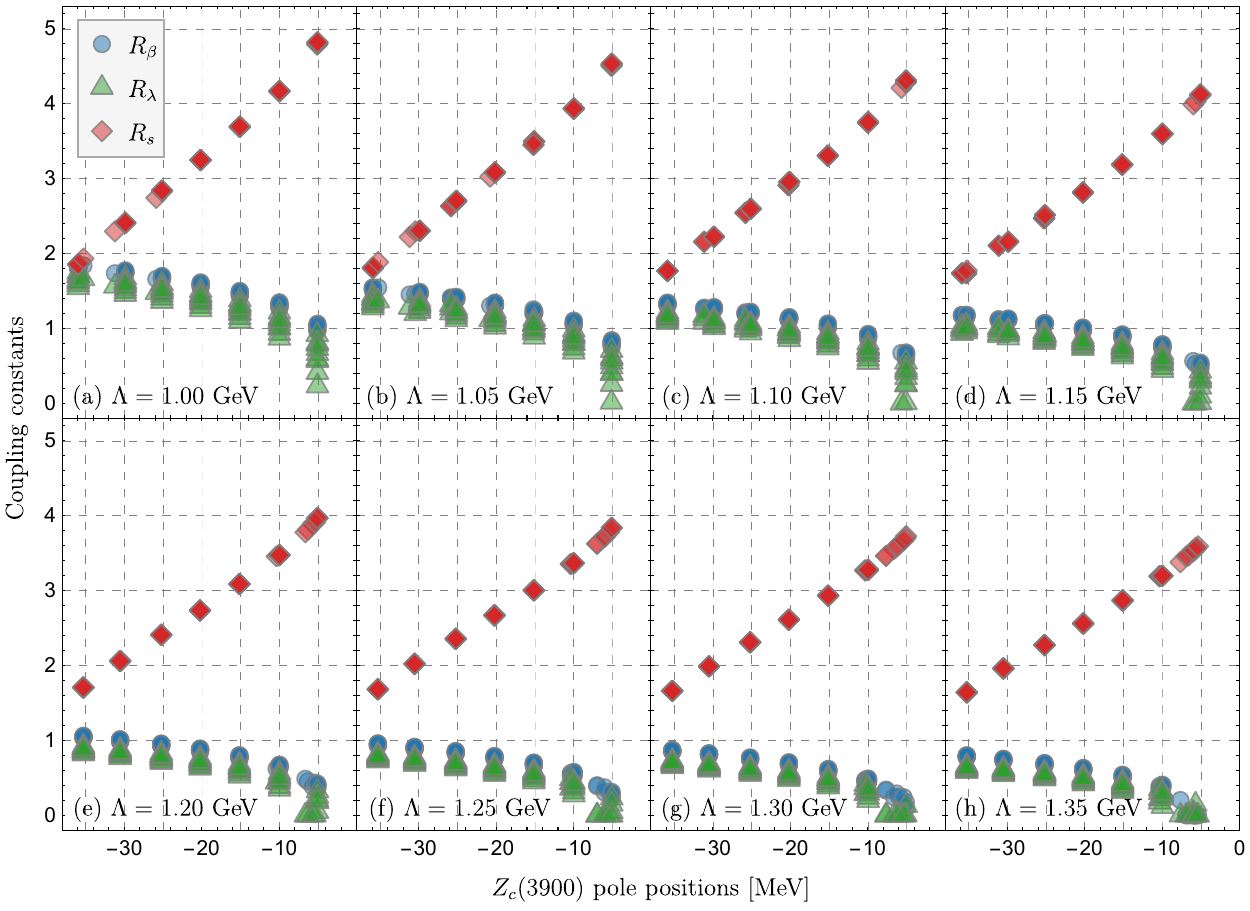}
    \caption{The coupling constants for the scalar- and vector-meson-exchange interactions determined by the pole positions of $X(3872)$, $T_{cc}(3875)$, and $Z_c(3900)$, as specified in Eq.~\eqref{eq:pole-position}. The ratios $R_\beta$, $R_\lambda$, and $R_s$ are defined relative to the baseline values, as detailed in Eq.~\eqref{eq:coupling-ratio} and Table~\ref{tab:lag_para}. Variations at the same \(Z_c\) pole position arise from different fitting parameters for the \(T_{cc}(3875)\) and \(X(3872)\) pole positions. }
    \label{fig:coupling}
\end{figure*}

To address the cutoff dependence, we vary the cutoff from 1.00 GeV to 1.35 GeV in steps of 0.05 GeV. For each cutoff and each pole set, we determine a corresponding set of $R_\lambda$, $R_\beta$, and $R_s$. The resulting coupling constants for different cutoffs are shown in Fig.~\ref{fig:coupling}. It is worth noting that among the three poles, the $Z_c(3900)$ exhibits the largest variation in its position. Therefore, we focus on the coupling constants associated with the $Z_c(3900)$ poles. During the fitting process, for most combinations of input parameters, only a single solution was obtained. However, for a very small number of combinations of cutoff and pole position, no real solutions were found for $R_\lambda$, $R_\beta$, and $R_s$. This occurs because these factors must be positive and real. In such cases, we provide the closest achievable results to the target poles. As shown in Fig.~\ref{fig:coupling} and Fig.~\ref{fig:bound}, some points are not perfectly aligned with the horizontal lines spaced at 5 MeV intervals from -35 to -5 MeV. This indicates a slight deviation between the fitted position of \(Z_{c}(3900)\) and the target value. Despite these deviations, the data still effectively capture our primary objective: establishing the relationship between \(Z_{c}(3900)\) and the \(DDD^*\) system.
From Fig.~\ref{fig:coupling}, it is evident that the scalar meson coupling constant varies significantly, ranging from 2 to 5 times the baseline values in Table~\ref{tab:lag_para} (model-I). This implies that the strength of the scalar meson potential varies between 4 and 25 times the baseline. For $Z_c(3900)$, the $\rho$- and $\omega$-exchange interactions almost cancel each other out, rendering the $Z_c(3900)$ pole primarily sensitive to $R_s$. The substantial uncertainties in the $Z_c(3900)$ pole position contribute to a wide range of possible scalar meson coupling values. 

The scalar exchange interaction is attractive as shown in Fig.~\ref{fig:V-baseline}, meaning that a larger scalar meson coupling constant moves the $Z_c(3900)$ virtual state pole closer to the threshold. The strong scalar-meson coupling in model-II, with $R_s \approx 4.5$, corresponds to a near-threshold $Z_c(3900)$ pole. In contrast, the weaker scalar coupling in the model-I results in a pole position farther than -35 MeV from the threshold. Notably, the $Z_c(3900)$ pole positions, which are reported to be several tens of MeV below the threshold in Refs.~\cite{Albaladejo:2015lob, Nakamura:2023obk, Yu:2024sqv}, qualitatively support the scenario of weaker scalar coupling. In comparison to the scalar-meson-exchange interaction, the variation in the vector-meson exchanges is much smaller, remaining consistently around 1. This indicates that the baseline choice for vector-meson exchange is relatively reasonable.

\begin{figure*}
    \centering
    \includegraphics[width=0.85\textwidth]{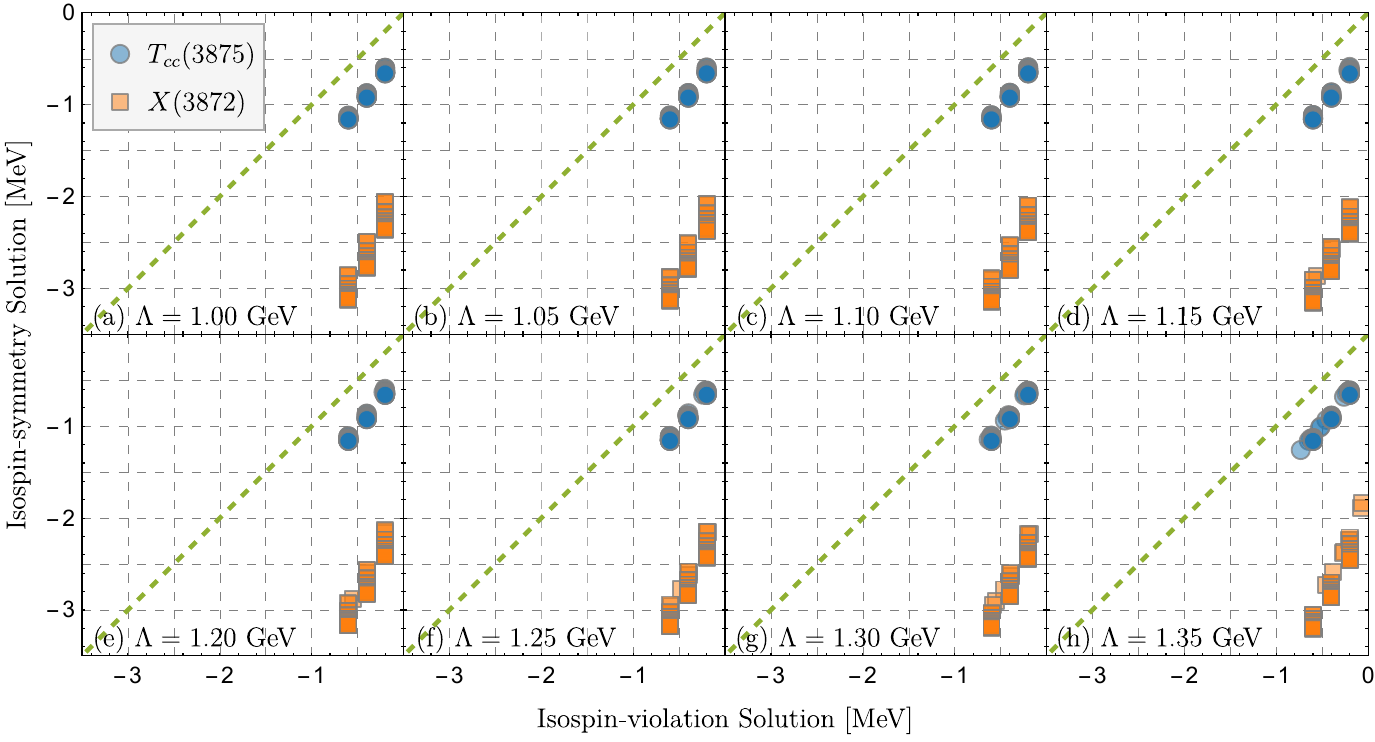}
    \caption{Comparison of the pole positions of $T_{cc}(3875)$ and $X(3872)$ in isospin-violating scenario  and the isospin limit scenario.}
    \label{fig:XT-iso}
\end{figure*}

The coupling constants determined in this work can be accessed via Zenodo repository~\cite{zhu_2024_14464767}. These parameters can be utilized to investigate other systems. Additionally, we have calculated the pole positions of $T_{cc}(3875)$ and $X(3872)$ in the isospin symmetry limit scenario and compared them with those in the isospin-violating scenario, as shown in Fig.~\ref{fig:XT-iso}. It can be observed that $X(3872)$ and $T_{cc}(3875)$, with binding energies around 0.2-0.6 MeV, would become bound states with binding energies around 2-3.5 MeV and 0.5-1.5 MeV, respectively. Apparently, the isospin violation effect appears to play a more significant role in \( X(3872) \) than in \( T_{cc}(3875) \).

\section{$DDD^*$ three-body systems}~\label{sec:3-body}
\subsection{Methodology}
The three-body $DDD^*$ system was studied using the nonrelativistic Hamiltonian given by:
\begin{equation}
H=\sum_{i=1}^3\left(m_i+\frac{p_i^2}{2m_i}\right)+\sum_{i<j}V_{ij}.
\end{equation}
where $m_i$ and $p_i$ are the mass and momentum of \(D/D^*\). $V_{ij}$ is the pair interaction of OBE which has been discussed in  Sec.~\ref{sec:obe}. In Ref.~\cite{Wu:2021kbu}, the isospin violation effects from the mass splitting and Coulomb interactions were included. However, the uncertainties of the three-body systems mainly arise from the indefinite strong interaction even in the isospin symmetry limit. Therefore, in this work, we will not consider the minor effect from the isospin violation and Coulomb interaction.

The total wave function of the $DDD^*$ system can be expressed as the direct product of the spatial wave function $\phi$, the isospin wave function $\chi_f$, and the spin wave function $\chi_s$:  
\begin{eqnarray}
\Psi = \mathcal{S}\left(\phi \otimes \chi_f \otimes \chi_s\right), \label{eq:wavefunction}
\end{eqnarray}
where \(\mathcal{S} = 1 + P_{ij}\) is the symmetrization operator, which ensures the wave function is symmetric under the exchange of two identical \(D\) mesons. Here, \(P_{ij}\) is the exchange operator. During the symmetrization process, the wave functions in different spaces are symmetrized simultaneously.  
In our approach, there is no exchange symmetry constraint imposed before applying the \(\mathcal{S}\) operator, which differs from the strategy used in Ref.~\cite{Wu:2021kbu}. An advantage of our method is that the wave function space is more general, and wave functions that do not conform to exchange symmetry are automatically excluded. See Ref.~\cite{Meng:2023jqk} for further details. In this work, we focus solely on the ground-state solution, where the total angular momentum \(L = 0\) is assumed for the spatial component. The \(D\) meson has a spin of 0, while the \(D^*\) meson has a spin of 1, resulting in a total spin of 1 for the system. The isospin of either the \(D\) or \(D^*\) meson is \(1/2\), which allows the total isospin of the system to be either \(1/2\) or \(3/2\). Both cases are investigated in this study.  For the isospin-\(3/2\) system, there is only one channel: two particles form an isospin triplet, which then combines with the third particle to create an \(I = 3/2\) system.  The isospin-\(1/2\) system, however, consists of two possible isospin channels. Using the state \(|I, I_3\rangle = |\frac{1}{2}, \frac{1}{2}\rangle\) as an example, the isospin wave functions are given by:  
\begin{eqnarray}
\left|\frac{1}{2}, \frac{1}{2}\right\rangle : \left\{
\begin{array}{l}
\sqrt{\frac{2}{3}} D^+ D^+ D^{*0} - \sqrt{\frac{1}{6}} \big(D^+ D^0 + D^0 D^+\big) D^{*+}, \\
\sqrt{\frac{1}{2}} \big(D^+ D^0 - D^0 D^+\big) D^{*+},
\end{array}
\right.    \label{eq:isowav}
\end{eqnarray}
where the two \(D\) mesons can form either an isospin singlet or triplet. Each configuration can then combine with the \(D^*\) to produce a total isospin-\(1/2\) system. 
It is important to note that other isospin bases can be chosen, as long as the isospin wave functions form a complete set. These bases are equivalent in describing the system~\cite{Meng:2023jqk}.

\begin{figure}[]    \includegraphics[width=0.45\textwidth]{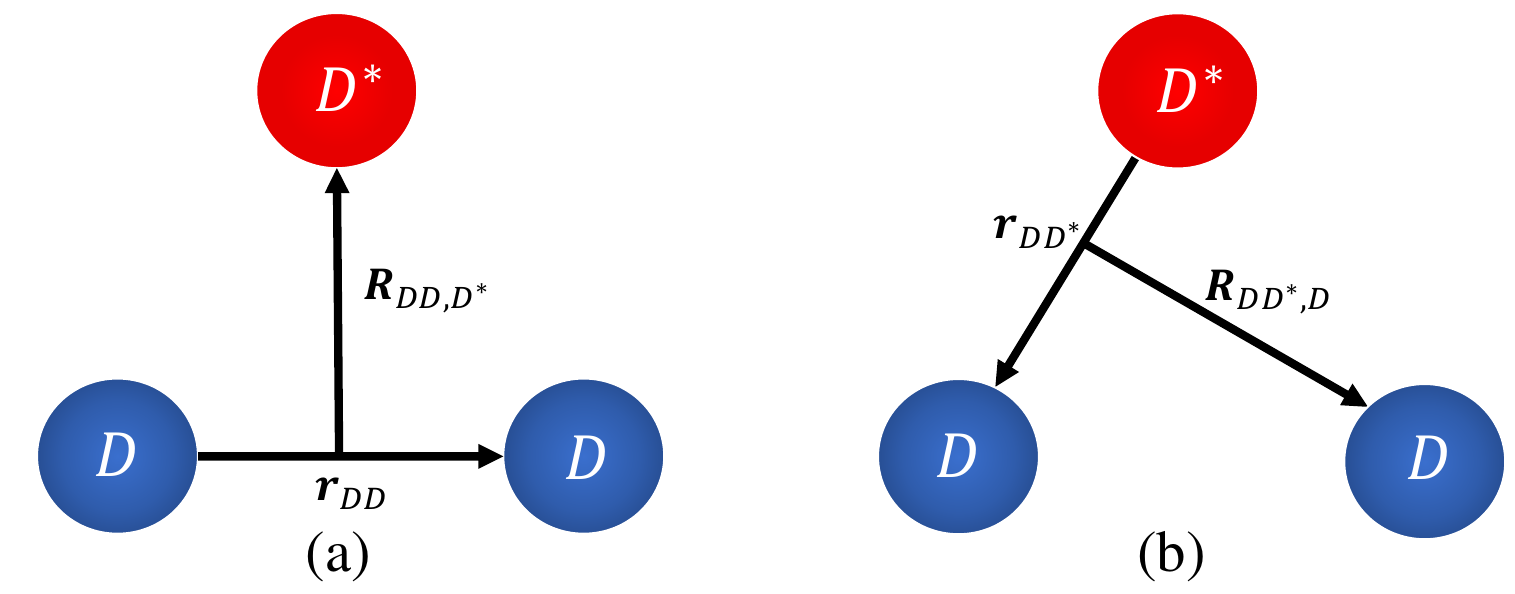}
    \caption{The two sets of Jacobi coordinates corresponding to different spatial configurations, where blue disks indicate two identical bosonic $D$ mesons  and red disk the $D^*$ meson. The third set of Jocobi coordinates by exchanging two $D$ mesons in (b) is not shown, however, it is included in the calculation. }
    \label{fig:jac}
\end{figure}

The spatial wave function is addressed in GEM, which is widely employed in few-body calculations and has been demonstrated to be a highly effective technique~\cite{Hiyama:2003cu}. We employ Jacobi coordinates as shown in Fig.~\ref{fig:jac} and work in the center-of-mass frame to eliminate the motion of the center of mass. In GEM, each wave function of a Jacobi coordinate is expanded  with the following basis:
\begin{equation}
\phi_{nlm}(\boldsymbol{r})=\sqrt{\frac{2^{l+5/2}}{\Gamma\left(l+\frac{3}{2}\right)r_n^3}}\left(\frac{r}{r_n}\right)^le^{-\frac{r^2}{r_n^2}}Y_{lm}(\hat{r}),
\label{eq:spacialwv}
\end{equation}
where \( Y_{lm} \) represents the spherical harmonic function, and the Gaussian size parameter \( r_n \) is taken in the form of geometric progression, $r_n=r_1a^{n-1}$. 
If we only use one single set of Jacobi coordinates, the wave function completeness requires an infinite number of Gaussian wave functions including orbital excited basis functions. However, it has been demonstrated that a finite set of Gaussian basis functions in a geometric progression, covering both long-range and short-range correlations, provides an efficient approximation of the general radial part of the wave functions~\cite{Hiyama:2003cu}. For the angular momentum component, since the ground-state solutions are primarily dominated by S-wave functions, we employ S-wave Gaussian basis functions in all relevant Jacobi coordinates of the system. Additionally, including S-wave functions from all Jacobi coordinates compensates for contributions from higher partial-wave angular momentum components in a single coordinate to some extent. Therefore, we adopt a finite set of Gaussian bases in all Jacobi coordinates to achieve rapid convergence of the results. In this work, the corresponding parameter values of bases are:
\begin{equation}
   \left\{
\begin{array}
{lll}r^{AB}_1=0.015\ \mathrm{fm},&r^{AB}_{n_{\max}}=5\ \mathrm{fm},&n_{\max}=16 \\
R^{AB,C}_1=0.07\ \mathrm{fm},&R^{AB,C}_{n_{\max}}=10\ \mathrm{fm},&n_{\max}=16 \\
\end{array}\right.,
\label{eq:jacobiset}
\end{equation}
where $A,B,C$ represent $D$ or $D^*$ meson.

In the GEM, the wave function of the $DDD^*$ system with total angular momentum $J$ and isospin $I$, can be expressed as:
\begin{equation}
\Psi^{IJ}=\mathcal{S}\sum_{\mathrm{jac}}\sum_{\alpha,n_i}C_{\alpha,n_i}^{\mathrm{(jac)}}\left[\chi_f^{I_\alpha}\left[\chi_s^S\phi_{n_1,n_2}^{\mathrm{(jac)}}\right]^J\right].~\label{eq:wvIJ}
\end{equation}
The notation $\mathrm{(jac)}=$ (a), (b) denotes the two Jacobi coordinate channels (See Fig.~\ref{fig:jac}). \( \alpha \) specifies the isospin channel (See Eq.~\eqref{eq:isowav} for an example), \( n_1 \) and \( n_2 \) correspond to the parameters of the Jacobi basis sets in Eq.~\eqref{eq:jacobiset}, and $C_{\alpha,n_i}^{\mathrm{(jac)}}$ denotes the expansion coefficients, which are determined using the Rayleigh-Ritz variational method. \( J \) and \( I \) represent the total spin and total isospin respectively. $\phi_{n_1,n_2}^{\mathrm{(jac)}}$ is total spacial wave function defined as
\begin{equation}
\phi_{n_1,n_2}^{\mathrm{(jac)}}=\phi_{n_1}(r^\mathrm{jac})\phi_{n_2}(R^\mathrm{jac}),
\end{equation}
where \( r^\mathrm{jac} \) and \( R^\mathrm{jac} \) represent two independent Jacobi coordinates in $\mathrm{(jac)}$. 
The expression for Gaussian function \( \phi \) is given by Eq.~\eqref{eq:spacialwv}.  Notably, before the symmetrization only two sets of Jacobi coordinates are adopted. However, after  symmetrization, the third set by exchanging two $D$ mesons in set (b) will be included automatically.

Based on the form of the constructed total wave function, the Schrödinger equation,
\begin{equation}
    H\Psi^{IJ}=E\Psi^{IJ},
\end{equation}
can be reformulated as a generalized eigenvalue problem using the Rayleigh-Ritz variational method,
\begin{equation}
    \sum_j\left[H_{ij}-EN_{ij}\right]C_{j}=0,\label{eq:scho_matrix_eq}
\end{equation}
where $H_{ij}$ is the matrix element of Hamiltonian, and $N_{ij}$ is the normalization matrix element, $i,j$ are indices from $\{\mathrm{jac},\alpha,n_i\}$.

The GEM performs the expansion in a set of square-integrable wave functions, which is efficient for the bound state solution. However, it cannot be used directly to probe the resonant state with finite lifetime, since its wave function is not square-integrable. However, by using the CSM~\cite{Aguilar:1971ve,Aoyama:2006hrz,Balslev:1971vb}, rescaling the Schr\"odinger equation with a phase factor, the resonant state solution with finite lifetime can also be determined within GEM.
By introducing a rotation to the momentum $p$ and the radial coordinate $r$ in the Hamiltonian, namely:
\begin{eqnarray}
&&U(\theta)\boldsymbol{r}=\boldsymbol{r}e^{i\theta},\quad U(\theta)\boldsymbol{p}=\boldsymbol{p}e^{-i\theta},\\
&&H(\theta)=\sum_{i=1}^3(m_i+\frac{p_i^2e^{-2i\theta}}{2m_i})+\sum_{i<j}V_{ij}(r_{ij}e^{i\theta}),
\end{eqnarray}
thereby resonant states can be treated in a manner similar to bound states as shown in Eqs.~\eqref{eq:wvIJ}-\eqref{eq:scho_matrix_eq}. The results are presented in the complex energy plane, where bound states appear on the negative real axis, and continuum states lie along rays extending from various energy thresholds. A resonance with mass \( M_R \) and width \( \gamma_R \) is located at \( E_R = M_R - i\gamma_R/2 \) in the complex energy plane, where \( R \) denotes the resonance. The resonance can be solved when \( \theta > |\mathrm{Arg}(E_R)/2| \). As the rotation angle \( \theta \) varies, the positions of both bound states and resonant states remain unchanged, whereas continuum states rotate clockwise by an angle of \( 2\theta \). This property is utilized to distinguish resonant states from continuum states. When employing the CSM, the rotation angle \( \theta \) is chosen to be less than \( \pi/4 \).

The root-mean-square (RMS) radius serves as a significant physical quantity for characterizing the spatial structure of the studied system. Moreover, it is widely utilized to determine whether a multi-quark system corresponds to a molecular state or a compact state~\cite{Wu:2024euj,Wu:2024zbx}. For a bound state, the RMS radius is:
\begin{equation}    r_{ij}^{\mathrm{rms}}\equiv\sqrt{\frac{\left\langle\Psi^{IJ}\left|r_{ij}^2\right|\Psi^{IJ}\right\rangle}{\left\langle\Psi^{IJ}\mid\Psi^{IJ}\right\rangle}}.
\end{equation}
For the resonant state in CSM, one can introduce the RMS radius
\begin{equation}
    r_{ij}^{\mathrm{rms}}\equiv\mathrm{Re}\left[\sqrt{\frac{\left(\Psi^{IJ}(\theta)\left|r_{ij}^2e^{2i\theta}\right|\Psi^{IJ}(\theta)\right)}{\left(\Psi^{IJ}(\theta)\mid\Psi^{IJ}(\theta)\right)}}\right],
\end{equation}
where the  c-product ~\cite{Romo:1968tcz} in the following form is introduced to ensure the analyticity of the integral: 
\begin{equation}
    (\Psi_i\mid\Psi_j)\equiv\int\Psi_i(\boldsymbol{r})\Psi_j(\boldsymbol{r})\mathrm{d}^3\boldsymbol{r}.
\end{equation}

\begin{figure*}
    \centering
    \includegraphics[width=0.90\textwidth]{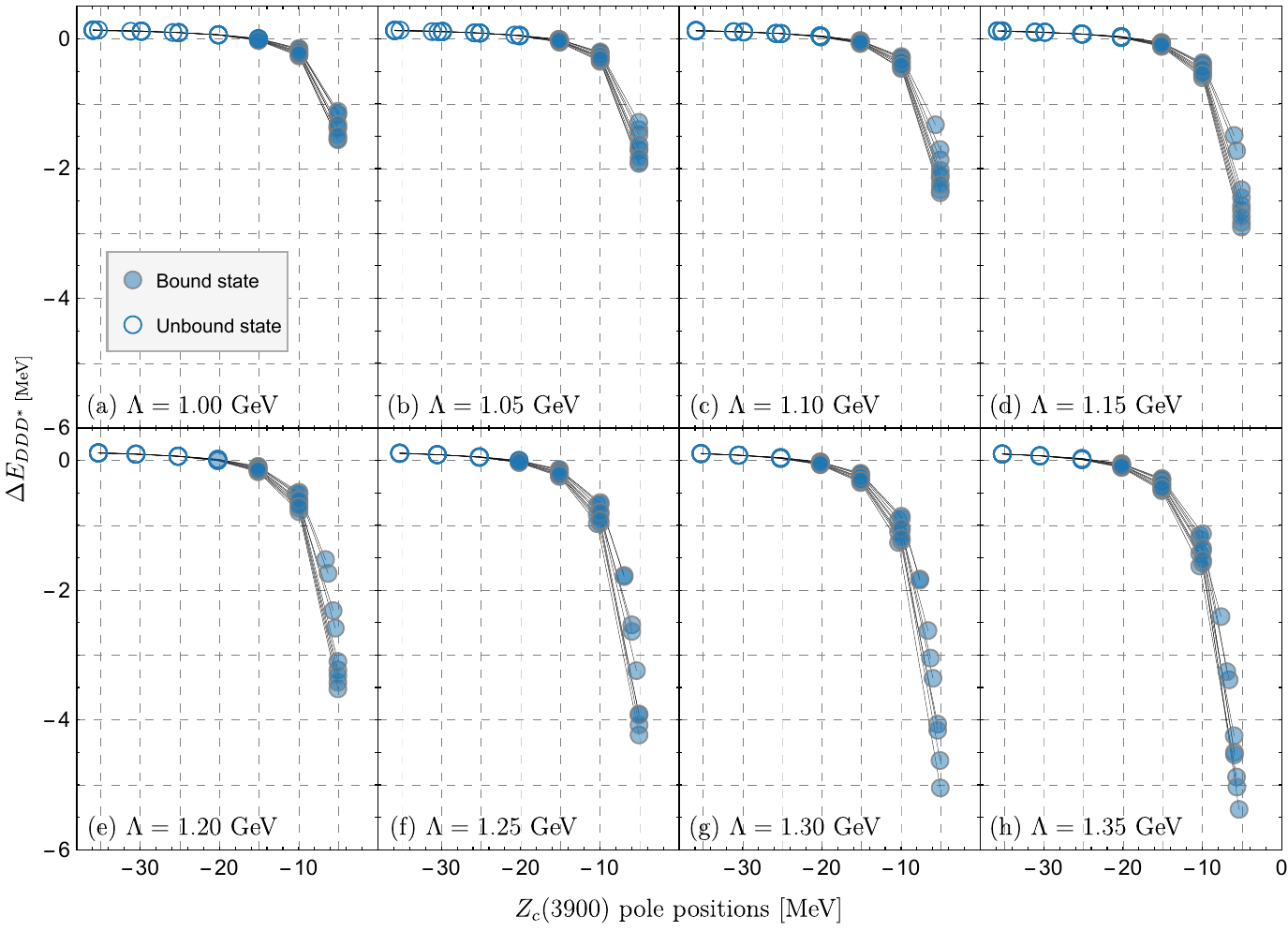}
    \caption{Variation of the three-body ($I = 1/2$) binding energy \(\Delta E_{DDD^*}\) with the \(Z_c(3900)\) pole position. \(\Delta E_{DDD^*}\) is measured relative to the \(T_{cc}D\) threshold, while the \(Z_c(3900)\) pole position is defined relative to the \(\DDbar\) threshold. Filled circles indicate bound states (\(\Delta E_{DDD^*} < 0\)), and hollow circles represent unbound states (\(\Delta E_{DDD^*} > 0\)). Variations at the same \(Z_c\)(3900) pole position arise from different fitting parameters for the \(T_{cc}(3875)\) and \(X(3872)\) pole positions.}
    
    \label{fig:bound}
\end{figure*}

\begin{figure*}
    \centering
    \includegraphics[width=0.95\textwidth]{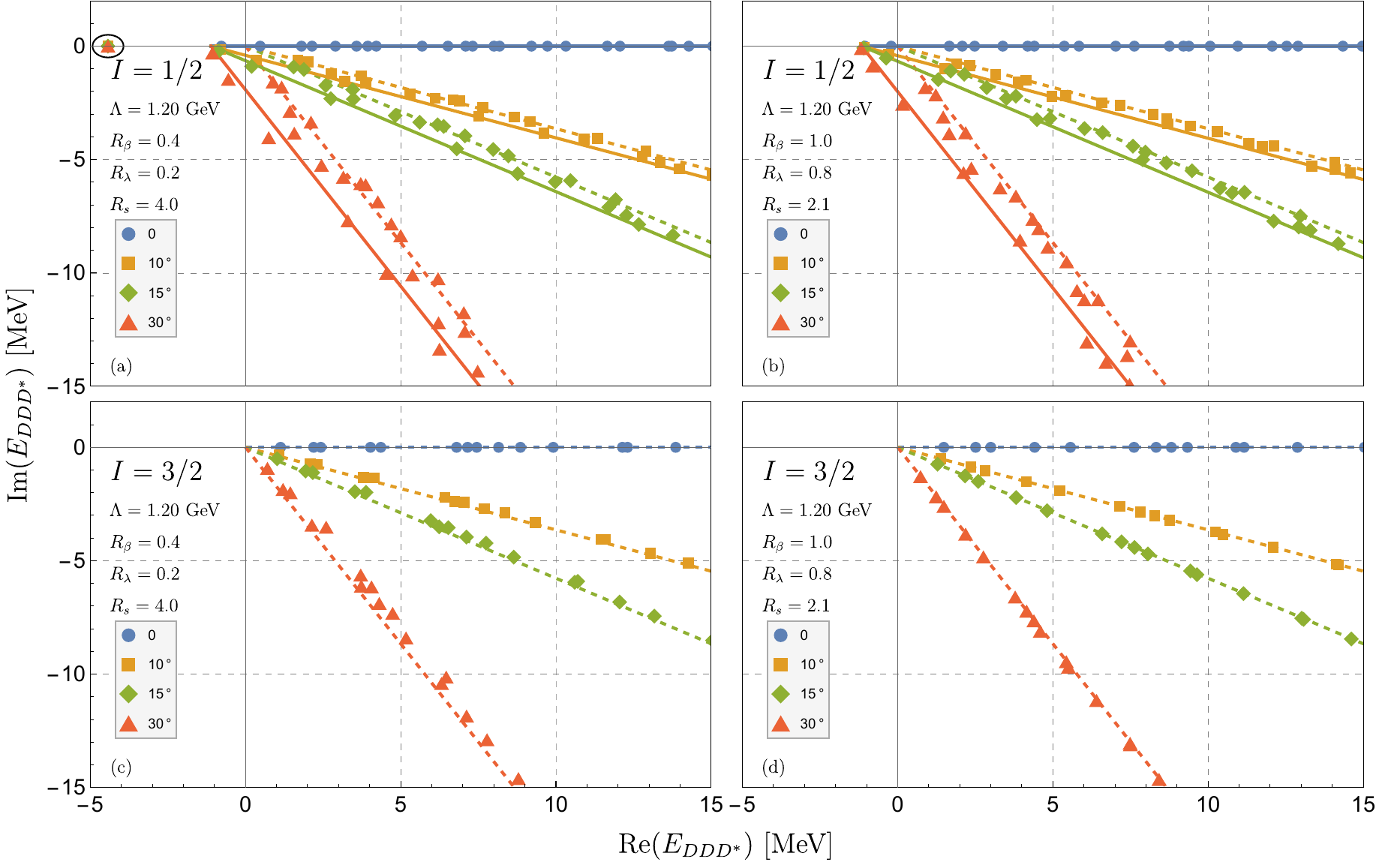}
    \caption{Variation of the three-body energy \( E_{DDD^*} \)  with the angle \( \theta \) of the CSM for the \( I = 1/2 \) and \( I = 3/2 \) channels with a cutoff of 1.20 GeV. The subfigures (a) and (c) are examples of the strong scalar-meson exchange while (b) and (d) are the examples of the weak scalar-meson exchange.  The solid rays and dashed rays start from the \( T_{cc} D\) threshold and the three-body \( DDD^* \) threshold, respectively. Three-body bound states are marked by black circles.}
    \label{fig:CSM}
\end{figure*}

\subsection{Numerical results and discussions}

In our three-body calculations, neither the isospin-breaking effects nor the channel-coupling effects involving configurations like \( D^*D^*D \) and \( D^*D^*D^* \) are considered. The binding energy of the isospin-\( \frac{1}{2} \) three-body system is defined relative to the \( T_{cc}D \) threshold, where \( T_{cc} \) is treated as a single-channel \( DD^* \) bound state in the isospin limit. The systematic uncertainties due to isospin violation and channel-coupling effects are expected to largely cancel out, provided that two-body and three-body calculations are performed in a consistent manner.

In the calculation of the $DDD^*$ system, we first employ the coupling constants in model-II~\cite{Liu:2019stu}, yielding consistent results with Ref.~\cite{Wu:2021kbu} and confirming the presence of a bound state in the parameter set. However, when using the parameters in model-I~\cite{Li:2012cs,Li:2012ss}, no bound state is observed, indicating that the existence of a bound state is strongly parameter-dependent. Comparing the two parameter sets given in Table~\ref{tab:lag_para}, it is apparent that the strengths of the scalar-meson-exchange interaction differ by approximately 20 times, which plays a significant role. The differences in the pseudoscalar-meson- and vector-meson-exchange strength are tiny. Additionally, another distinction is that the model-II does not include the $\eta$ exchange, whereas it is considered in model-I. However, since the $\eta$-exchange interaction is weak, it is not a key point.  

To investigate the possible variation of the $DDD^*$, we employ the refined parameters recalibrated by the \( X(3872) \), \( T_{cc}(3875) \), and \( Z_{c}(3900) \) pole positions in Sec.~\ref{sec:obe}. The results for $I=1/2$ system are shown in Fig.~\ref{fig:bound}, focusing on the connection between the $DDD^*$ system and \( Z_c(3900) \). One can see the existence of the three-body bound state is closely tied to the pole position of the \( Z_c(3900) \): As the \( Z_c(3900) \) pole position moves toward its threshold, the three-body system's binding energy increases, while the \( Z_c(3900) \) pole moves away from its threshold, the three-body binding energy decreases until no bound state exists. This connection can be easily understood through the variation in the strength of the \(\sigma\)-meson-exchange interaction. A near-threshold \( Z_c(3900) \) pole corresponds to a strong attractive \(\sigma\)-meson-exchange interaction, which tends to form a bound state. Conversely, a distant \( Z_c(3900) \) pole corresponds to a weak attractive \(\sigma\)-meson-exchange interaction, resulting in an unbound three-body system. In Fig.~\ref{fig:CSM}, examples of strong \(\sigma\)-exchange (\(R_s=4.0\)) and weak \(\sigma\)-exchange (\(R_s=2.1\)) interactions are shown in subfigures (a) and (b), respectively. Both subfigures exhibit the continuous spectrum of the \( T_{cc}D \) system, indicating the existence of two-body $DD^*$ bound state. However, only the strong coupling case in subfigure (a) demonstrates the presence of \( DDD^* \) bound states. It is unsurprising that the two parameter sets in Table~\ref{tab:lag_para} yield different results for bound and unbound \( DDD^* \) states, as the \(\sigma\)-exchange interactions in model-II (model-I) are stronger (weaker) than the strong (weak) example illustrated in Fig.~\ref{fig:CSM}. To investigate the spatial structures of the possible $DDD^*$ bound states, the RMS radii of $DD^*$ and $DD$ are presented in Appendix~\ref{app:rms}.

In Fig.~\ref{fig:bound}, the small variations in \( E_{DDD^*} \) at the same \( Z_c \) pole position result from different fitting parameters for the \( T_{cc}(3875) \) and \( X(3872) \) pole positions. This indicates that the uncertainties in the \( T_{cc}(3875) \) and \( X(3872) \) pole positions have minor impact on the existence of the \( DDD^* \) state. In Fig.~\ref{fig:bound}, results under different cutoffs are shown. Notably, varying  the cutoff would not alter the results qualitatively. Considering the uncertainties of  \( T_{cc}(3875) \) and \( X(3872) \) and the dependence on the cutoffs, the boundary between the bound and unbound $DDD^*$ systems is approximately located where the virtual state pole of \( Z_c \) lies between \(-15\) and \(-20\) MeV. The recent analyses in Refs.~\cite{Albaladejo:2015lob,Nakamura:2023obk,Yu:2024sqv} indicate the virtual state pole of \( Z_c(3900) \) is far below the threshold at least about 40 MeV. These analyses support the weak  $\sigma$-exchange scenario and the unbound $DDD^*$ conclusion.

In addition to the isospin-\( \frac{1}{2} \) \( DDD^* \) systems, the isospin-\( \frac{3}{2} \) system is also investigated. Using the parameters determined from the pole positions in Eq.~\eqref{eq:pole-position}, we find no evidence of isospin-1 \( DD^* \) or \( DD \) bound states, nor any isospin-\( \frac{3}{2} \) \( DDD^* \) bound states. Furthermore, we explore the potential existence of an isospin-\( \frac{3}{2} \) three-body resonance state using the complex scaling method but find no such resonance. Two representative results are presented in Fig.~\ref{fig:CSM}(c) and (d).

\section{conclusion}~\label{sec:concl}

The existence of the \(DDD^*\) bound state remains uncertain due to significant variations in the OBE interaction, particularly in the strength of scalar-meson-exchange interactions. We address this issue by constraining the bound \(DDD^*\) system using the \(Z_c(3900)\) pole positions, assuming it to be a virtual state. For the \(Z_c(3900)\) state, the pseudoscalar-meson coupling is well-determined, and the \(\rho\)- and \(\omega\)-exchange interactions nearly cancel each other out, leaving the coupling constant of the \(\sigma\)-exchange as the only unknown parameter. We solve the Schrödinger equation for the \(DDD^*\) three-body system via the Gaussian expansion method. The results reveal that the isospin-\(I=1/2\) $DDD^*$ bound state exists when the virtual state of \(Z_c(3900)\) in the \(DD^*/\bar{D}\bar{D}^*\) system lies near its threshold, within approximately \(-15\) MeV. However, the three-body system ceases to be bound when the \(Z_c(3900)\) pole is shifted farther away from the threshold, beyond about \(-20\) MeV.

To pin down or rule out the existence of the $DDD^*$ three-body bound state requires determining the strength of scalar-meson-exchange interaction in higher precision. However, it can be linked to the \( Z_c(3900) \) state from another perspective. If the three-body bound state is observed experimentally, the position of the \( Z_c(3900) \) pole can be inferred. Conversely, if the position of the \( Z_c(3900) \) pole is measured, the existence or nonexistence of the three-body bound state can be determined. Experimentally, the search for this three-body bound state can be pursued through the decay channels \( DDD^* \to DDD\pi \) and \( DDD^* \to DDD\gamma \)~\cite{Wu:2021kbu}. 

Additionally, the potential existence of \(DDD^*\) three-body resonances in isospin-1/2 and isospin-3/2 channels is explored using a combination of the Gaussian expansion method and the complex scaling method. However, no resonance state for the \(DDD^*\) system was found in any of the channels. 

As an another significant outcome, we recalibrate the coupling constants in the OBE model based on the pole positions of \(X(3872)\), \(T_{cc}(3875)\), and \(Z_c(3900)\), rigorously addressing the dependence on the cutoff. Eight sets of coupling constants corresponding to cutoff values ranging from 1.00 to 1.35 GeV in steps of 0.05 GeV are provided. These parameters provide a valuable resource for more accurate calculations of few-body systems involving \(D\), \(D^*\), and their antiparticles.

	\section*{ACKNOWLEDGMENTS}
	
	We thank Zi-Yang Lin and Jian-Bo Cheng for helpful discussions. This project was supported by the National Natural Science Foundation of China (Grant No. 12475137 and No. 12175318), ERC NuclearTheory (Grant No. 885150), and the Natural Science Foundation of Guangdong Province of China (Grant No. 2023A15150117 and No. 2022A1515011922). The computational resources were supported by High-performance Computing Platform of Peking University.

    \section*{Data Availability Statement}

The data supporting the findings of this study is available on Zenodo repository~\cite{zhu_2024_14464767} or can be obtained directly from the authors upon reasonable request.

	\appendix
	\section{Fourier transformation}\label{app:fourier}
In this section, we adopt the following normalization conventions:   
    \begin{eqnarray}
\langle\bm{r}|\bm{r}'\rangle&=&\delta(\bm{r}-\bm{r}'),\nonumber\\
\langle\bm{p}|\bm{p}'\rangle&=&(2\pi)^{3}\delta(\bm{p}-\bm{p}').
\end{eqnarray}
Thus, the plane wave function is given by $\langle\bm{r}|\bm{p}\rangle=e^{i\bm{p\cdot r}}$. The general nonlocal potential in coordinate space can be related to the potential in momentum space via Fourier transformation~\cite{Hoshizaki:1960emm}: 
    \begin{eqnarray}        V(\bm{r}',\bm{r})&=&\langle\bm{r}'|\hat{V}|\bm{r}\rangle \nonumber\\
    &=&\int\frac{d^{3}\bm{p}}{(2\pi)^{3}}\frac{d^{3}\bm{p'}}{(2\pi)^{3}}\langle\bm{r}'|\bm{p}'\rangle\langle\bm{p}'|V|\bm{p}\rangle\langle\bm{p}|\bm{r}\rangle \nonumber\\
    &=&\int\frac{d^{3}\bm{p}}{(2\pi)^{3}}\frac{d^{3}\bm{p'}}{(2\pi)^{3}}V(\bm{p}',\bm{p})e^{i(\bm{p'\cdot r'-p\cdot r})}\nonumber\\
    &=&\int\frac{d^{3}(\bm{k}/2)}{(2\pi)^{3}}\frac{d^{3}\bm{q}}{(2\pi)^{3}}V(\bm{p}',\bm{p})e^{i(\bm{k\cdot s}/2+\bm{q}\cdot\bm{R})}.
    \end{eqnarray}
In the last step of the above derivation, we introduce new sets of coordinates and momenta: 
\begin{equation}
    \begin{cases}
\bm{R}=\frac{\bm{r}+\bm{r}'}{2}, & \bm{q}=\bm{p}'-\bm{p},\\
\bm{s}=\bm{r'-r}, & \frac{\bm{k}}{2}=\frac{\bm{p}'+\bm{p}}{2}.
\end{cases}
\end{equation}
When $V(\bm{p}',\bm{p}) = \tilde{V}(\bm{q})$ depends only on the momentum $\bm{q}$, we obtain:
\begin{eqnarray}
V(\bm{r}',\bm{r})&=&\tilde{V}(\bm{R})\delta(\bm{s}),\nonumber\\
\tilde{V}(\bm{R})&=&\int\frac{d^{3}\bm{q}}{(2\pi)^{3}}e^{i\bm{q}\cdot\bm{R}}\tilde{V}(\bm{q}).\label{eq:FTq}
\end{eqnarray}
The coordinate space representation of the potential acting on the wave function is:
\begin{eqnarray}
    \langle\bm{r}|\hat{V}|\psi\rangle&=&\int d^{3}\bm{r}'V(\bm{r},\bm{r}')\psi(\bm{r}')\nonumber\\
    &=&\int d^{3}\bm{r}'\tilde{V}\left(\frac{\bm{r}+\bm{r}'}{2}\right)\delta(\bm{r}-\bm{r}')\psi(\bm{r}')\nonumber\\
    &=&\tilde{V}(\bm{r})\psi(\bm{r}).
\end{eqnarray}
Therefore, we obtain the results for local interactions. When $V(\bm{p}’,\bm{p}) = \tilde{V}(\bm{k})$ depends only on $\bm{k}$, the potential in coordinate space is:
\begin{eqnarray}
    V(\bm{r}',\bm{r})&=&\int\frac{d^{3}\bm{k}}{(2\pi)^{3}}\frac{d^{3}(\bm{q}/2)}{(2\pi)^{3}}\tilde{V}(\bm{k})e^{i(\bm{k\cdot s}/2+\bm{q}\cdot\bm{R})}\nonumber\\&=&\tilde{V}\left(\frac{\bm{s}}{2}\right)\delta(2\bm{R}),
\end{eqnarray}
with 
\begin{eqnarray}
    \tilde{V}\left(\frac{\bm{s}}{2}\right)=\int\frac{d^{3}\bm{k}}{(2\pi)^{3}}e^{i\bm{k\cdot s}/2}\tilde{V}(\bm{k}).\label{eq:FTk}
\end{eqnarray}
Therefore, the potential acting on a wave function becomes: 
\begin{eqnarray}
    \langle\bm{r}|\hat{V}|\psi\rangle&=&\int d^{3}\bm{r}'\tilde{V}\left(\frac{\bm{r}-\bm{r}'}{2}\right)\delta\left(\frac{\bm{r}+\bm{r}'}{2}\right)\psi(\bm{r}')\nonumber\\&=&\tilde{V}(\bm{r})\psi(-\bm{r}).
\end{eqnarray}
 It is interesting to note the additional sign that appears for the argument of the wave function compared to the local potential. If $\psi(\bm{r})$ contains only even partial wave components, then $\psi(-\bm{r}) = \psi(\bm{r})$. Conversely, for odd partial wave states, $\psi(-\bm{r}) = -\psi(\bm{r})$.

For the specific potential in this work, the Fourier transformation reads: 
\begin{eqnarray}
    \frac{1}{u^{2}+\bm{q}^{2}}F(u,\Lambda,q^{2})^{2}&\to&H_{0}(u,\Lambda,r),\nonumber\\
    \frac{\bm{q}^{2}}{u^{2}+\bm{q}^{2}}F(u,\Lambda,q^{2})^{2}&\to&-H_{1}(u,\Lambda,r),\\
    \frac{q_{i}q_{j}}{u^{2}+\bm{q}^{2}}F(u,\Lambda,q^{2})^{2}&\to&-[H_{3}(u,\Lambda,r)T_{ij}+H_{1}(u,\Lambda,r)\frac{\delta_{ij}}{3}],\nonumber
\end{eqnarray}
where $T_{ij} = \frac{3r_{i}r_{j}}{r^{2}} - \delta_{ij}$. One can also easily obtain the transformation for the case where $V(\bm{p’},\bm{p}) = \tilde{V}(\bm{k})$. The explicit expressions of $H_0$, $H_1$, and $H_3$ are: 
\begin{eqnarray}
  H_{0}(u,\Lambda,r)&=&\frac{u}{4\pi}\left[\frac{e^{-ur}-e^{-\Lambda r}}{ur}-\frac{\Lambda^{2}-u^{2}}{2u\Lambda}e^{-\Lambda r}\right],\nonumber\\
  H_{1}(u,\Lambda,r)&=&\frac{u^{3}}{4\pi}\left[\frac{e^{-ur}-e^{-\Lambda r}}{ur}-\frac{(\Lambda^{2}-u^{2})\Lambda^{2}}{2u^{3}\Lambda}e^{-\Lambda r}\right],\nonumber\\
  H_{3}(u,\Lambda,r)&=&\frac{u^{3}}{12\pi}\left[ -\frac{e^{-\Lambda r}\Lambda^{2}\left(\frac{3}{\Lambda^{2}r^{2}}+\frac{3}{\Lambda r}+1\right)}{ru^{3}}\right.\nonumber\\&&
  ~~~~~~~~~~~-\frac{e^{-\Lambda r}(\Lambda r+1)\left(\Lambda^{2}-u^{2}\right)}{2ru^{3}}\nonumber\\&&
  ~~~~~~~~~~~\left.+\frac{e^{-ur}\left(\frac{3}{r^{2}u^{2}}+\frac{3}{ru}+1\right)}{ru}\right].  \label{eq;HHH}
\end{eqnarray}

\section{Root-mean-square radii}~\label{app:rms}
\begin{figure*}
    \centering    \includegraphics[width=0.90\textwidth]{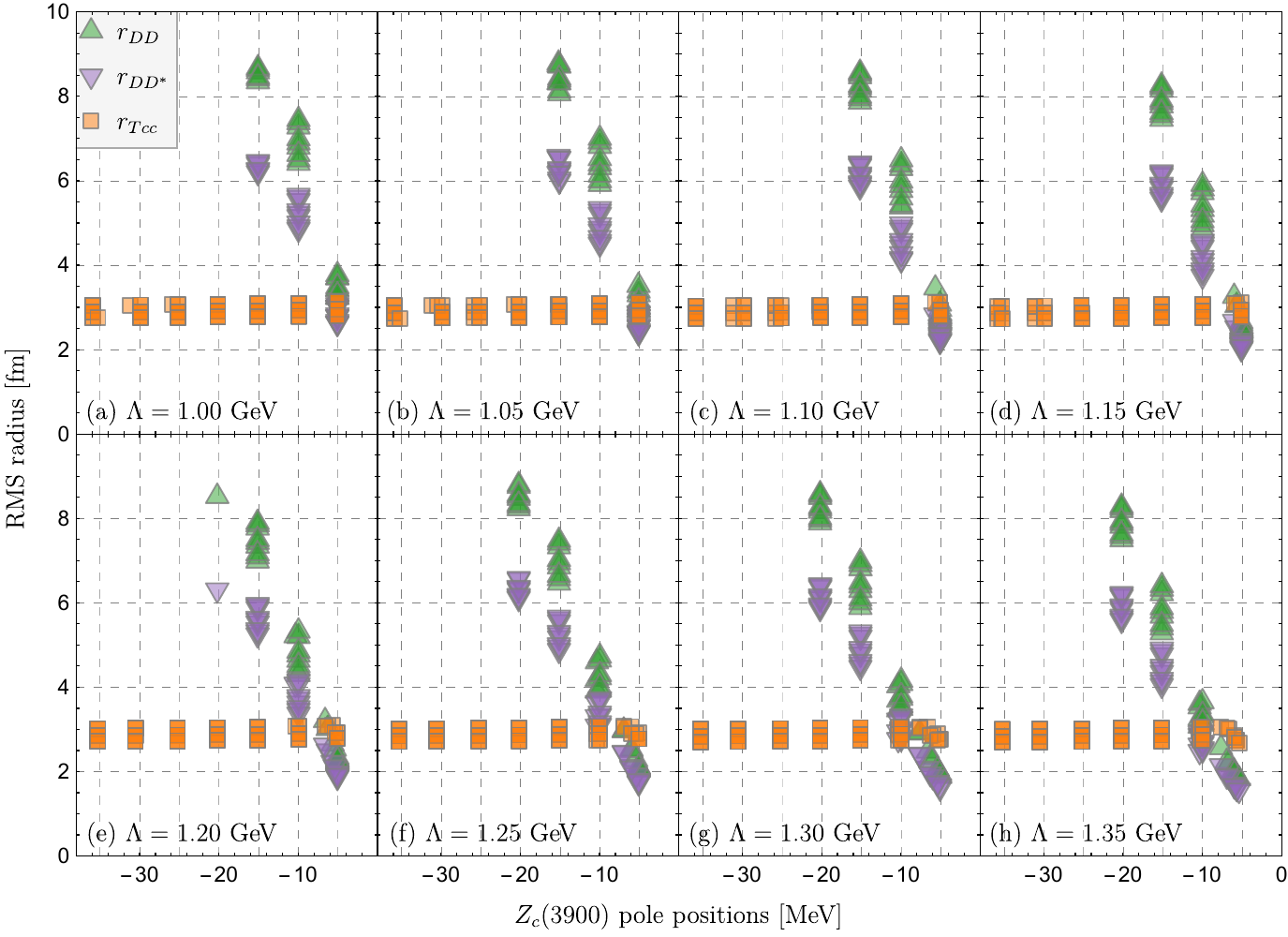}
    \caption{The relationship between the RMS radii of \( DD^* \), \( DD \)  in three-body bound state system \( DDD^* \), \( DD^* \) in two-body bound state system \( T_{cc}(3875) \) and the position of the \( Z_c(3900) \) pole, under eight different sets of cutoff \( \Lambda \).
    }
    \label{fig:RMS}
\end{figure*}

To investigate the spatial structures of the possible $DDD^*$ bound states, the RMS radii of $DD^*$ and $DD$ compared with that of $T_{cc}(3875)$ are presented in Fig.~\ref{fig:RMS}. It should be noticed that the RMS radii results are based on the Gaussian bases in Eq.~\eqref{eq:jacobiset}.

	\bibliography{ref}

\end{document}